\documentclass[showpacs,preprintnumbers,amsmath,amssymb 
]{revtex4}
\usepackage{graphicx}
\usepackage{dcolumn}
\usepackage{bm}

\begin{document}

\title{Transition to Zero Cosmological Constant and Phantom Dark
Energy\\
as Solutions Involving Change of Orientation of Space-Time
Manifold}

\author
{E. I. Guendelman \thanks{guendel@bgu.ac.il} and A.  B. Kaganovich
\thanks{alexk@bgu.ac.il}}
\address{Physics Department, Ben Gurion University of the Negev, Beer
Sheva 84105, Israel}

\date{\today}

\begin{abstract}
The main conclusion of long-standing discussions concerning the
role of solutions with degenerate metric ($g\equiv
det(g_{\mu\nu})=0$ and even with $g_{\mu\nu}=0$)  was that in the
first order formalism they are physically acceptable and must be
included in the path integral. In particular, they may describe
topology changes and reduction of "metrical dimension" of
space-time. The latter implies  disappearance of the volume
element $\sqrt{-g}d^4x$ of a 4-D space-time in a neighborhood of
the point with $g=0$. We pay attention that besides $\sqrt{-g}$,
the 4-D space-time differentiable manifold possesses also a
"manifold volume measure" (MVM) described by a 4-form which is
sign indefinite and generically independent of the metric.  The
first order formalism proceeds with originally independent
connection and metric structures of the space-time manifold. In
this paper we bring up the question whether the first order
formalism should be supplemented with degrees of freedom of the
space-time differentiable manifold itself, e.g. by means of the
MVM. It turns out that adding the MVM degrees of freedom to the
action principle in the first order formalism one can realize very
interesting dynamics.  Such Two Measures Field Theory enables
radically new approaches to resolution of the cosmological
constant problem. We show that fine tuning free solutions
describing a transition to $\Lambda =0$ state involve oscillations
of $g_{\mu\nu}$ and MVM around zero. The latter can be treated as
a dynamics involving changes of orientation of the space-time
manifold. As we have shown earlier, in realistic scale invariant
models (SIM), solutions formulated in the Einstein frame satisfy
all existing tests of General Relativity (GR). Here we reveal
surprisingly that in SIM, all ground state solutions with
$\Lambda\neq 0$ appear to be degenerate either in $g_{00}$ or  in
MVM. Sign indefiniteness of MVM in a natural way yields a
dynamical realization of a phantom cosmology ($w<-1$). It is very
important that for all solutions, the metric tensor rewritten in
the Einstein frame has regularity properties exactly as in GR. We
discuss  new physical effects which arise from this theory and in
particular strong gravity effect in high energy physics
experiments.

\end{abstract}

\maketitle

\section{Introduction: Degenerate Metric, Manifold Volume Measure and  Orientation
of the Space-Time Manifold}

Solutions with degenerate metric were a subject of a long-standing
discussions starting probably with the paper by Einstein and
Rosen\cite{Einstein}. In spite of some difficulty interpreting
solutions with degenerate metric in classical theory of
gravitation, the prevailing view was that they have physical
meaning and must be included in the path
integral\cite{Hawking1979},\cite{D'Auria-Regge},\cite{Tseytlin1982}.
As it was shown in Refs.\cite{Hawking1979},\cite{Horowitz}, in the
first order formulation of an appropriately extended general
relativity, solutions with $g(x)\equiv \det(g_{\mu\nu})=0$ can
describe changes of the space-time topology. Similar idea is
realized also in the Ashtekar's
variables\cite{Ashtekar-3-in-Jacobson},\cite{Jacobson-2-in-Jacobson}.
There are known also classical
solutions\cite{Dray-0}-\cite{Senovilla} with change of the
signature of the metric tensor.

 The space-time regions with $g(x)=0$
can be treated  as having {\em 'metrical dimension'} $D<4$ (using
terminology by Tseytlin\cite{Tseytlin1982}).

The simplest solution with $g(x)=0$ is $g_{\mu\nu}=0$ while the
affine connection is arbitrary (or, in the Einstein-Cartan
formulation, the vierbein $e_{\mu}^a =0$ and $\omega_{\mu}^{ab}$
is arbitrary). Such  solutions have been studied by D'Auria and
Regge\cite{D'Auria-Regge}, Tseytlin\cite{Tseytlin1982}),
Witten\cite{Witten-15-and-16-in-Horowitz},
Horowitz\cite{Horowitz}, Giddings\cite{Giddings},
Ba\~{n}ados\cite{Banados}; it has been suggested that
$g_{\mu\nu}=0$ should be interpreted as essentially non-classical
phase in which diffeomorphism invariance is unbroken and it is
realized at high temperature and curvature.

Now we would like to bring up a question: whether the equality
$g(x)=0$ really with a necessity means that the dimension of the
space-time manifold in a small neighborhood of the point $x$ may
become $D<4$? At first sight it should be so because the volume
element is
\begin{equation}
 dV_{(metrical)}=\sqrt{-g}d^{4}x.
 \label{dVg}
\end{equation}
 Note that the latter is the "metrical" volume
 element, and the possibility to describe the volume of the space-time manifold
 in this way appears after the 4-dimensional differentiable manifold $M_4$ is equipped with the
  metric structure. For a solution with $g_{\mu\nu} =0$, the situation with description of the space-time
   becomes even worse .
    However, in spite of lack of the metric, the manifold
 $M_4$ may still  possess a nonzero volume element and have the dimension $D=4$. The well
  known way to realize it consists
 in the construction of a differential 4-form build for example
 by means of four differential 1-forms $d\varphi_a$, ($a=1,2,3,4$): \,
 $d\varphi_1\wedge d\varphi_2\wedge d\varphi_3\wedge d\varphi_4$.
 Each of the 1-forms $d\varphi_a$ may be defined by a scalar field
  $\varphi_a(x)$. The appropriate volume element of the 4-dimensional
  differentiable manifold $M_4$
 can be represented in the following way
\begin{equation}
dV_{(manifold)} = {4!}d\varphi_{1}\wedge d\varphi_{2}\wedge
d\varphi_{3}\wedge d\varphi_{4}\equiv \Phi d^{4}x \label{dV}
\end{equation}
where
\begin{equation}
\Phi \equiv \varepsilon_{abcd} \varepsilon^{\mu\nu\lambda\sigma}
(\partial_{\mu}\varphi_{a}) (\partial_{\nu}\varphi_{b})
(\partial_{\lambda}\varphi_{c}) (\partial_{\sigma}\varphi_{d}).
\label{Phi}
\end{equation}
is the volume measure independent of $g_{\mu\nu}$ as opposed to
the case of the metrical volume measure $\sqrt{-g}$. In order to
emphasize the fact that the volume element (\ref{dV}) is metric
independent we will call it a {\em manifold volume element} and
the measure $\Phi$ - a {\em manifold volume measure}.

If $\Phi(x)\neq 0$ one can think of four scalar fields
$\varphi_a(x)$ as describing a  homeomorphism of an open
neighborhood of the point $x$ on the 4-dimensional Euclidean space
$R^4$. However if one allows a dynamical mechanism of metrical
dimensional reduction of the space-time by means of degeneracy of
the metrical volume measure $\sqrt{-g}$, there is no reason to
ignore a possibility of a similar effect permitting degenerate
manifold volume measure $\Phi$. The possibility of such (or even
stronger, with a sign change of $\Phi$) dynamical effect seems to
be here more natural since {\em the manifold volume measure $\Phi$
is sign indefinite} (in  Measure Theory, sign indefinite measures
are known as signed measures\cite{signed}) . Note that the
metrical and manifold volume measures are not obliged generically
to be simultaneously nonzero.

The original idea to use differential forms as describing
dynamical degrees of freedom of the space-time differentiable
manifold has been developed by Taylor in his attempt\cite{Taylor}
to quantize the gravity. Taylor argued that quantum mechanics is
not compatible with a Riemannian metric space-time; moreover, in
the quantum regime space-time is not even an affine manifold. Only
in the classical limit the metric and connection emerge, that one
allows then to construct a traditional space-time description. Of
course, the transition to the classical limit is described  in
Ref.\cite{Taylor} rather in the form of a general prescription.
Thereupon we would like to pay attention to the additional
possibility which was ignored in Ref.\cite{Taylor}. Namely, in the
classical limit not only the metric and connection emerge but also
some of the differential forms  could keep (or restore) certain
dynamical effect in the classical limit. In such a case, the
traditional space-time description may occur to be incomplete.
{\bf Our key idea} is that one of such lost differential forms,
the 4-form (\ref{dV}) survives in the classical limit as
describing dynamical degrees of freedom of the volume measure of
the space-time manifold, and hence can affect the gravity theory
on the classical level too\footnote{An opposite view on the role
of the volume element has been studied by Wilczek\cite{Wilczek}.
Another idea of modified volume element was studied in
Ref.\cite{Mosna}.}.

If we add four scalar fields $\varphi_a(x)$ as new  variables to a
set of usual variables  (like metric, connection and matter
degrees of freedom) which undergo variations in the action
principle\footnote{Fore a more detailed discussion of the  role of
scalars $\varphi_a(x)$ in the TMT dynamics, see the end of
Sec.II.} then one can expect an effect of gravity and matter on
the manifold volume measure $\Phi$ and vice versa. We will see
later in this paper that in fact such effects exist and in
particular classical cosmology solutions of a significant interest
exist where $\Phi$ vanishes and changes sign.

As is well known, the 4-dimensional differentiable manifold is
 orientable if it possesses a differential form  of degree 4
which is nonzero at every point on the manifold. Therefore two
possible signs of the manifold volume measure (\ref{Phi}) are
associated with two possible orientations of the space-time
manifold.
 The latter means that besides a
dimensional reduction and topology changes on the level of the
differentiable manifold, the incorporation of the manifold volume
measure $\Phi$ allows to realize solutions describing {\em
dynamical change of the orientation of the space-time manifold}.

In the light of existence of two volume measures, the simplest way
 to take into account this fact in the action principle
consists in  the modification of the action which should now
consist of two terms, one with the usual measure $\sqrt{-g}$ and
another - with the measure $\Phi$,
\begin{equation}
    S_{mod} = \int \left(\Phi L_1 +\sqrt{-g}L_2\right) d^{4}x,
\label{sec-2-S-modif}
\end{equation}
where two Lagrangians $L_1$ and $L_2$ coupled with manifold and
metrical volume measures appear respectively.  According to our
previous experience\cite{GK1}-\cite{GK10} in Two Measures Field
Theory (TMT) we will proceed with an additional basic assumption
that, at least on the classical level, the Lagrangians $L_1$ and
$L_2$ are independent of the scalar fields $\varphi_a(x)$, i.e.
the manifold volume measure degrees of freedom enter into TMT only
through the manifold volume measure $\Phi$.  In such a case,
  the action (\ref{sec-2-S-modif}) possesses
an infinite dimensional symmetry
\begin{equation}
\varphi_{a}\rightarrow\varphi_{a}+f_{a}(L_{1}), \label{IDS}
\end{equation}
 where
$f_{a}(L_{1})$ are arbitrary functions of  $L_{1}$ (see details in
Ref.\cite{GK3}). One can hope that this symmetry should prevent
emergence of the scalar fields $\varphi_a(x)$ dependence in
$L_{1}$ and $L_{2}$ after quantum effects are taken into account.

Note that Eq.(\ref{sec-2-S-modif}) is just a convenient way for
presentation of the theory in a general form. In concrete models
studied in the present paper, we will see that the action
(\ref{sec-2-S-modif}) can be always rewritten in an equivalent
form where {\em each term in the action has its own  total volume
measure and the latter is a linear combination of $\Phi$ and
$\sqrt{-g}$}.

In the next section it will be shown that  the space-time geometry
described in terms of the original metric and connection of the
underlying action (\ref{sec-2-S-modif}) is not generically
Riemannian. However by making use a change of variables to the
Einstein frame one can represent the resulting equations of motion
in the Riemannian (or Einstein-Cartan) space-time.

In our previous investigations  we have shown that TMT enables
radically new approaches to resolution of the cosmological
constant (CC) problem\cite{GK3},\cite{G1},\cite{GK9} (for an
alternative approach see Ref.\cite{Comelli}). Intrinsic features
of TMT allow to realize a scalar field dark energy model where all
dependence of the scalar field appears as a result of spontaneous
breakdown of the dilatation symmetry. Solutions of this model
formulated in the Einstein frame satisfy all existing tests of
General Relativity (GR)\cite{GK6},\cite{GK7},\cite{GK10}. A new
sort of dynamical protection from the initial singularity of the
curvature becomes possible\cite{GK9}. It also allows us to realize
a phantom dark energy in the late time universe without explicit
introducing phantom scalar field\cite{GK9}.

 In contrast to all our previous investigations of
TMT, {\em the purpose of the present paper} is to study the
dynamics of the metric $g_{\mu\nu}$ and the manifold volume
measure $\Phi$ (used in the underlying action
(\ref{sec-2-S-modif})) in a number of TMT models. The main
attention is concentrated on the analysis of the amazing features
of "irregularity" of $g_{\mu\nu}$ and $\Phi$ (involving change of
orientation of the space-time manifold) in the course of
transitions to a ground state and in the phantom dark energy. It
is very important to note immediately that in the Einstein frame
the metric tensor has regularity properties exactly as in GR. The
organization of the paper is the following.  In Sec. II we discuss
general features of classical dynamics in TMT. In Sec. III we
consider the pure gravity model. In Secs. IV and V, in the
framework of a simple scalar field model I, we analyze in detail
the behavior of two volume measures in the course of transition to
the ground state with zero CC. In Sec. VI we study a (generically
broken) intrinsic TMT symmetry which is restored in the ground
states; the relation of this symmetry restoration to the old CC
problem\cite{Weinberg1} is also analyzed; a discussion of this
effect is continued in Sec.VIII. In Sec.VII we shortly present the
scalar field model II with spontaneously broken global scale
invariance\cite{G1} studied in detail in Ref.\cite{GK9}. In the
framework of such class of models, an interesting dynamics of the
metric and the manifold volume measure in the course of transition
to ground states is analyzed in Sec.VIII. In section IX we reveal
that  a possibility to realize a phantom dark energy without
explicit introducing a phantom scalar field (demonstrated in
\cite{GK9}) has the origin in a sign indefiniteness of the
manifold volume measure (\ref{Phi}). Finally, since one cannot
check directly whether a tiny/zero cosmological constant is
fine-tuned or not, in Sec.X we discuss the new physical effects
which arise from this theory and in particular a strong gravity
effect in high energy physics experiments.

\section{Classical equations of motion}
 Varying the measure fields $\varphi_{a}$, we get
$B^{\mu}_{a}\partial_{\mu}L_{1}=0$ where
$B^{\mu}_{a}=\varepsilon^{\mu\nu\alpha\beta}\varepsilon_{abcd}
\partial_{\nu}\varphi_{b}\partial_{\alpha}\varphi_{c}
\partial_{\beta}\varphi_{d}$.
Since $Det (B^{\mu}_{a}) = \frac{4^{-4}}{4!}\Phi^{3}$ it follows
that if $\Phi\neq 0$ the constraint
\begin{equation}
 L_{1}=sM^{4} =const.
\label{varphi}
\end{equation}
must be satisfied, where $s=\pm 1$ and $M$ is a constant of
integration with the dimension of mass. Variation of the metric
$g^{\mu\nu}$ gives
\begin{equation}
\zeta\frac{\partial L_1}{\partial g^{\mu\nu}}+\frac{\partial
L_2}{\partial g^{\mu\nu}}-\frac{1}{2}g_{\mu\nu}L_2 =0,
\label{g-mu-nu-varying}
\end{equation}
where
\begin{equation}
 \zeta\equiv \frac{\Phi}{\sqrt{-g}}
 \label{zeta}
\end{equation}
 is the scalar field
build of the scalar densities $\Phi$ and $\sqrt{-g}$.

We study  models with  the Lagrangians of the form
\begin{equation}
L_1=-\frac{1}{b_g\kappa}R(\Gamma, g)+L_1^m, \quad
L_2=-\frac{1}{\kappa}R(\Gamma, g)+L_2^m \label{L1L2}
\end{equation}
 where $\Gamma$ stands
for affine connection, $R(\Gamma,
g)=g^{\mu\nu}R_{\mu\nu}(\Gamma)$,
$R_{\mu\nu}(\Gamma)=R^{\lambda}_{\mu\nu\lambda}(\Gamma)$ and
$R^{\lambda}_{\mu\nu\sigma}(\Gamma)\equiv \Gamma^{\lambda}_{\mu\nu
,\sigma}+ \Gamma^{\lambda}_{\alpha\sigma}\Gamma^{\alpha}_{\mu\nu}-
(\nu\leftrightarrow\sigma)$. Dimensionless factor $b_g^{-1}$ in
front of $R(\Gamma, g)$ in $L_1$ appears because there is no
reason for couplings  of the scalar curvature to the measures
$\Phi$ and $\sqrt{-g}$ to be equal. We choose $b_g>0$ and $\kappa
=16\pi G$, $G$ is the Newton constant. $L_1^m$ and $L_2^m$ are the
matter Lagrangians which can include all possible terms used in
regular (with only volume measure $\sqrt{-g}$) field theory
models.

Since the measure $\Phi$ is sign indefinite, the total volume
measure $(\Phi/b_g +\sqrt{-g})$ in the gravitational term
$-\kappa^{-1}\int R(\Gamma, g)(\Phi/b_g +\sqrt{-g})d^4x$  is
generically also sign indefinite.

Variation of the connection yields the equations we have solved
earlier\cite{GK3}. The result is
\begin{equation}
\Gamma^{\lambda}_{\mu\nu}=\{
^{\lambda}_{\mu\nu}\}+\frac{1}{2}(\delta^{\alpha}_{\mu}\sigma,_{\nu}
+\delta^{\alpha}_{\nu}\sigma,_{\mu}-
\sigma,_{\beta}g_{\mu\nu}g^{\alpha\beta}) \label{GAM2}
\end{equation}
where $\{ ^{\lambda}_{\mu\nu}\}$  are the Christoffel's connection
coefficients of the metric $g_{\mu\nu}$ and
\begin{equation}
\sigma,_{\mu}\equiv \frac{1}{\zeta
+b_g}\zeta,_{\mu},\label{sigma-mu}
\end{equation}

If $\zeta\neq const.$ the covariant derivative of $g_{\mu\nu}$
with this connection is nonzero (nonmetricity) and consequently
geometry of the space-time with the metric $g_{\mu\nu}$ is
generically non-Riemannian.  The gravity and matter field
equations obtained by means of the first order formalism contain
both $\zeta$ and its gradient as well. It turns out that at least
at the classical level, the measure fields $\varphi_a$ affect the
theory only through the scalar field $\zeta$.

For the class of models (\ref{L1L2}), the consistency of the
constraint (\ref{varphi}) and the gravitational equations
(\ref{g-mu-nu-varying}) has the form of the following  constraint
\begin{equation}
 (\zeta
-b_g)(sM^4-L_1^m)+g^{\mu\nu}\left(\zeta \frac{\partial
L_{1m}}{\partial g^{\mu\nu}}+\frac{\partial L_2^m}{\partial
g^{\mu\nu}}\right)-2L_2^m=0, \label{Constr-original}
\end{equation}
which determines  $\zeta(x)$ (up to the chosen value of the
integration constant $sM^4$) as a local function of matter fields
and metric. Note that the geometrical object $\zeta(x)$ does not
have  its own dynamical equation of motion and its space-time
behavior is totally determined by  the metric and matter fields
dynamics via the constraint (\ref{Constr-original}). Together with
this, since $\zeta$ enters into all equations of motion, it
generically has straightforward effects on  dynamics of the matter
and gravity through the forms of potentials, variable fermion
masses and selfinteractions\cite{GK1}-\cite{GK10}.

For understanding the structure of TMT it is important to  note
that TMT (where, as we suppose, the scalar fields $\varphi_a$
enter only via the measure $\Phi$) is a constrained dynamical
system. In fact, the volume measure $\Phi$ depends only upon the
first derivatives of fields $\varphi_a$ and this dependence is
linear.  The  fields $\varphi_a$ do not have their own dynamical
equations: they are auxiliary fields.  All their dynamical effect
is displayed only in the following two ways: a) in generating the
constraint (\ref{varphi}) (or (\ref{Constr-original}));  b) in the
appearance of the scalar field $\zeta$ and its gradient in all
equations of motion.

\section{Pure Gravity TMT model }

Let us start from the simplest TMT model with action
(\ref{sec-2-S-modif}) where
\begin{equation}
L_1=-\frac{1}{b_g\kappa}R(\Gamma, g)-\Lambda_1, \quad
L_2=-\frac{1}{\kappa}R(\Gamma, g)-\Lambda_2 \label{L1L2pure}
\end{equation}
and $\Lambda_1$, $\Lambda_2$ are constants. Note that $\Lambda_1
=const.$ cannot have a physical contribution to the field
equations (in this model - only gravitational) because $\Phi
\Lambda_1$ is a total derivative. Nevertheless we keep $\Lambda_1$
to see explicitly how $\Lambda_1$ appears in the result.
$\Lambda_2/2$ would have a sense of the cosmological constant in
the regular, non TMT, theory (i.e.  with the only measure
$\sqrt{-g}$).

Following the procedure described in Sec.II we obtain the
gravitational equations (\ref{g-mu-nu-varying}) and the constraint
(\ref{Constr-original}) in the following form:
\begin{equation}
R_{\mu\nu}(\Gamma)=\frac{\kappa}{2}\frac{b_g\Lambda_2}{\zeta
-b_g}g_{\mu\nu}
 \label{Grav.eq.Pure}
\end{equation}
\begin{equation}
\zeta=b_g-\frac{2\Lambda_2}{sM^4+\Lambda_1}=const.,
\label{Constr.Pure}
\end{equation}
where $sM^4$ is the constant of integration that appears in
Eq.(\ref{varphi}) and we have assumed that the total volume
measure in the gravitational term of the action is nonzero, that
is $\Phi/b_g+\sqrt{-g}\neq 0$.

Since $\zeta =const.$ the connection $\Gamma^{\lambda}_{\mu\nu}$,
Eq.(\ref{GAM2}), coincides with the Christoffel's connection
coefficients of the metric $g_{\mu\nu}$. Therefore in the model
under consideration, the space-time with the metric $g_{\mu\nu}$
is (pseudo) Riemannian. It follows from Eqs.(\ref{Grav.eq.Pure})
and (\ref{Constr.Pure}) the resulting Einstein equations
\begin{equation}
G_{\mu\nu}(g)=\frac{\kappa}{2}\Lambda g_{\mu\nu}; \qquad \Lambda
=\frac{b_g}{2}(sM^4+\Lambda_1) \label{Ein.eq.Pure}
\end{equation}

Constancy of $\zeta(x)$ on the mass shell, Eq.(\ref{Constr.Pure}),
means that for the described solution the manifold and metrical
volume measures coincide up to a normalization factor. However,
this is true only on the mass shell; if we were try to start from
this assumption in the underlying action the resulting solution
would be different completely.

The model possesses a few interesting features in what it concerns
the CC:

(1) The effective CC $\Lambda$ appears as a constant of
integration (as we noticed above, the parameter $\Lambda_1$ has no
a physical meaning and it can be absorbed by the constant of
integration). The effective regular, non TMT, gravity theory
provides the same equations if the cosmological constant  is added
explicitly.

(2) The effective cosmological constant $\Lambda$ does not depend
at all on the CC-like parameter $\Lambda_2$ (which should describe
a total vacuum energy density including vacuum fluctuations of all
matter fields). The latter resembles the situation in the
unimodular theory\cite{unimodular-1},\cite{unimodular-2}.

(3) Note that  $\Lambda$ becomes very small if the integration
constant is chosen such that $sM^4+\Lambda_1$ is very small. The
latter is equivalent to a solution with $\Phi/b_g\gg \sqrt{-g}$.
In the limit where the metrical volume measure $\sqrt{-g}\to 0$
while the manifold volume measure $\Phi$ remains nonzero, we get
$\Lambda\to 0$. Thus a $\Lambda =0$ state is realized for a
solution which involves a reduction of the metrical dimension to
$D^{(g)}<4$ and at the same time the dimension of the space-time
as a differentiable manifold remains $D^{(m)}=4$.

(4) In the limit where the free parameter $b_g\to \infty$, the
 gravitational term in the underlying
action (Eqs.(\ref{sec-2-S-modif}) and (\ref{L1L2pure})) with
coupling to the manifold measure $\Phi$ approaches zero; then TMT
takes the form of a regular (non TMT) field theory, but the
effective cosmological constant $\Lambda$ becomes infinite. If we
wish to reach a very small value of $\Lambda$ keeping the
integration constant arbitrary, one should take the opposite limit
where $b_g\ll 1$. Then in the underlying action, the weight of the
gravitational term with coupling to the manifold volume measure
$\Phi$  increases with respect to the regular one with coupling to
the metrical measure $\sqrt{-g}$.

The above speculations can be regarded as a strong indication that
TMT possesses a  potential for resolution of the CC problem. In
the next sections we will study this issue in more realistic
models.

\section{Scalar Field Model I}

Let us now study a model including gravity as in Eqs.(\ref{L1L2})
and a scalar field $\phi$. The action has the same structure as in
Eq.(\ref{sec-2-S-modif}) but  it is more convenient to write down
it in the following form
\begin{equation}
S_{mod}^{(1)}=\frac{1}{b_g}\int d^4x \left[-\frac{1}{\kappa}(\Phi
+b_g\sqrt{-g})R(\Gamma,g)+(\Phi
+b_{\phi}\sqrt{-g})\frac{1}{2}g^{\mu\nu}\phi_{,\mu}\phi_{,\nu}
-\Phi V_1(\phi)-\sqrt{-g}\,V_2(\phi)\right]
\label{S-model-scalar.f.}
\end{equation}
The appearance of the dimensionless factor $b_{\phi}$ is explained
by the fact that without fine tuning it is impossible in general
to provide the same coupling of the $\phi$ kinetic term to the
measures $\Phi$ and $\sqrt{-g}$. $V_1(\phi)$ and $V_2(\phi)$ are
potential-like functions; we will see below that the physical
potential of the scalar $\phi$ is a complicated function of
$V_1(\phi)$ and $V_2(\phi)$.

The constraint (\ref{Constr-original}) reads now
\begin{equation}
(\zeta
-b_g)[sM^4+V_1(\phi)]+2V_2(\phi)+b_g\frac{\delta}{2}g^{\alpha\beta}\phi_{,\alpha}\phi_{,\beta}
=0, \label{Scalar-f-Constr-original}
\end{equation}
where
\begin{equation}
\delta =\frac{b_g-b_{\phi}}{b_g} \label{delta}
\end{equation}
Since $\zeta\neq const.$ the connection (\ref{GAM2}) differs from
the connection of the metric $g_{\mu\nu}$. Therefore the
space-time with the metric $g_{\mu\nu}$ is non-Riemannian. To see
the physical meaning of the model we perform a transition to a new
metric
\begin{equation}
\tilde{g}_{\mu\nu}=(\zeta +b_{g})g_{\mu\nu}, \label{gmunuEin}
\end{equation}
where the connection $\Gamma^{\lambda}_{\mu\nu}$ becomes equal to
the Christoffel connection coefficients of the metric
$\tilde{g}_{\mu\nu}$ and the space-time turns into (pseudo)
Riemannian. This is why the set of dynamical variables using the
metric $\tilde{g}_{\mu\nu}$ we call the Einstein frame. One should
point out that {\it the transformation} (\ref{gmunuEin}) {\em is
not a conformal} one since $(\zeta +b_{g})$ is sign indefinite.
But $\tilde{g}_{\mu\nu}$ is a regular pseudo-Riemannian metric.
For the action (\ref{S-model-scalar.f.}), gravitational equations
(\ref{g-mu-nu-varying}) in the Einstein frame take canonical GR
form with the same $\kappa =16\pi G$
\begin{equation}
G_{\mu\nu}(\tilde{g}_{\alpha\beta})=\frac{\kappa}{2}T_{\mu\nu}^{eff}
\label{gef}
\end{equation}
Here  $G_{\mu\nu}(\tilde{g}_{\alpha\beta})$ is the Einstein tensor
in the Riemannian space-time with the metric $\tilde{g}_{\mu\nu}$
and the energy-momentum tensor reads
\begin{eqnarray}
T_{\mu\nu}^{eff}&=&\frac{\zeta +b_{\phi}}{\zeta +b_{g}}
\left(\phi_{,\mu}\phi_{,\nu}- \tilde{g}_{\mu\nu}X\right)
-\tilde{g}_{\mu\nu}\frac{b_{g}-b_{\phi}}{(\zeta +b_{g})} X
+\tilde{g}_{\mu\nu}V_{eff}(\phi;\zeta,M)
 \label{Tmn-2}
\end{eqnarray}
where
\begin{equation}
X\equiv\frac{1}{2}\tilde{g}^{\alpha\beta}\phi_{,\alpha}\phi_{,\beta}
\label{X}
\end{equation}
and the function $V_{eff}(\phi;\zeta,M)$ is defined as following:
\begin{equation}
V_{eff}(\phi;\zeta ,M)= \frac{b_g\left[sM^{4}+V_{1}(\phi)\right]
-V_{2}(\phi)}{(\zeta +b_{g})^{2}}. \label{Veff1}
\end{equation}

The scalar $\phi$ field equation following from
Eq.(\ref{S-model-scalar.f.}) and rewritten in the Einstein frame
reads
\begin{equation}
\frac{1}{\sqrt{-\tilde{g}}}\partial_{\mu}\left[\frac{\zeta
+b_{\phi}}{\zeta
+b_{g}}\sqrt{-\tilde{g}}\tilde{g}^{\mu\nu}\partial_{\nu}\phi\right]
+\frac{\zeta V_1^{\prime}+ V_2^{\prime}}{(\zeta +b_{g})^2}=0
 \label{phief}
\end{equation}

The scalar field $\zeta$ in Eqs.(\ref{Tmn-2})-(\ref{phief}) is
determined by means of the consistency equation
(\ref{Constr-original}) which in the Einstein frame
(\ref{gmunuEin}) takes the form
\begin{equation}
(\zeta -b_{g})[sM^{4}+V_1(\phi)]+ 2V_2(\phi)+\delta\cdot
b_{g}(\zeta +b_{g})X =0.\label{constraint2-1}
\end{equation}

\section{Manifold Measure and Old Cosmological Constant Problem:
Cosmological Dynamics with $|\Phi|/\sqrt{-g}\to \infty$}

It is interesting to see the role of the manifold volume measure
in the resolution of the CC problem. We accomplish this now in the
framework of the scalar field model I of previous section. The
$\zeta$-dependence of $V_{eff}(\phi;\zeta ,M)$, Eq.(\ref{Veff1}),
in the form of inverse square like $(\zeta +b_{g})^{-2}$ has a key
role in the resolution of the old CC problem in TMT. One can show
that if quantum corrections to the underlying action generate
nonminimal coupling like  $\propto R(\Gamma,g)\phi^2$ in both
$L_1$ and $L_2$, the general form of the $\zeta$-dependence of
$V_{eff}$ remains similar: $V_{eff}\propto (\zeta +f(\phi))^{-2}$,
where $f(\phi)$ is a function. The fact that only such type of
$\zeta$-dependence emerges in $V_{eff}(\phi;\zeta ,M)$, and a
$\zeta$-dependence is absent for example in the numerator of
$V_{eff}(\phi;\zeta ,M)$, is a direct result of our basic
assumption that $L_1$ and $L_2$ in the action
(\ref{sec-2-S-modif}) are independent of the manifold measure
fields $\varphi_a$.

Generically, in the action (\ref{S-model-scalar.f.}),
$b_{\phi}\neq b_g$ that yields a nonlinear kinetic term (i.e. the
$k$-essence type dynamics) in the Einstein frame\footnote{See also
Ref.\cite{GK9}  where we study in detail a model with dilatation
symmetry which also results in the $k$-essence type dynamics}. But
for purposes of this section it is enough to take a simplified
model with $b_{\phi}= b_g$ (which is in fact a fine tuning) since
the nonlinear kinetic term has no qualitative effect on the zero
CC problem. In such a case solving the constraint
(\ref{constraint2-1}) for $\zeta$ and substituting into
Eqs.(\ref{Tmn-2})-(\ref{phief}) we obtain equations for
scalar-gravity system which can be described by the regular GR
effective action with the scalar field potential
\begin{equation}
 V_{eff}(\phi)=
\frac{(sM^{4}+V_{1})^2}{4[b_g(sM^4+V_1(\phi))-V_{2}(\phi)]}
\label{Veff3}
\end{equation}
For an arbitrary nonconstant function $V_1(\phi)$ there exist
infinitely many values of the integration constant $sM^4$ such
that  $V_{eff}(\phi)$ has the {\it absolute minimum} at some
$\phi=\phi_0$ with $V_{eff}(\phi_0)=0$ (provided
$b_g[sM^4+V_1(\phi)]-V_2(\phi)>0$). This effect takes place as
$sM^4+V_1(\phi_0)=0$ {\bf without fine tuning of the parameters
and initial conditions}. Note that the choice of the scalar field
potential in the GR effective action in a form proportional to a
perfect square like emerging in Eq.(\ref{Veff3}) would mean a fine
tuning.

 For illustrative purpose  let us consider the model
with
\begin{equation}
V_1(\phi)=\frac{1}{2}\mu_1^2\phi^2, \qquad
V_2(\phi)=V_2^{(0)}+\frac{1}{2}\mu_2^2\phi^2. \label{V12model}
\end{equation}
 Recall that adding a
constant to $V_1$ does not effect equations of motion, while
$V_2^{(0)}$ absorbs the bare CC and all possible vacuum
contributions. We take negative integration constant, i.e. $s=-1$,
and the only restriction on the values of the integration constant
$M$ and the parameters is that denominator in (\ref{Veff3}) is
positive\footnote{For the case $s=+1$ and the ground state with
nonzero cosmological constant see Appendix A}.

Consider spatially flat FRW universe with the metric in the
Einstein frame
\begin{equation}
\tilde{g}_{\mu\nu}=diag(1,-a^2,-a^2,-a^2), \label{FRW}
\end{equation}
 where $a=a(t)$ is the
scale factor. Each cosmological solution ends with the transition
to a $\Lambda =0$ state via damping oscillations of the scalar
field $\phi$ towards its  absolute minimum $\phi_0$. The
appropriate oscillatory regime in the  phase plane is presented in
Fig. 1.
\begin{figure}[htb]
\begin{center}
\includegraphics[width=10.0cm,height=6.0cm]{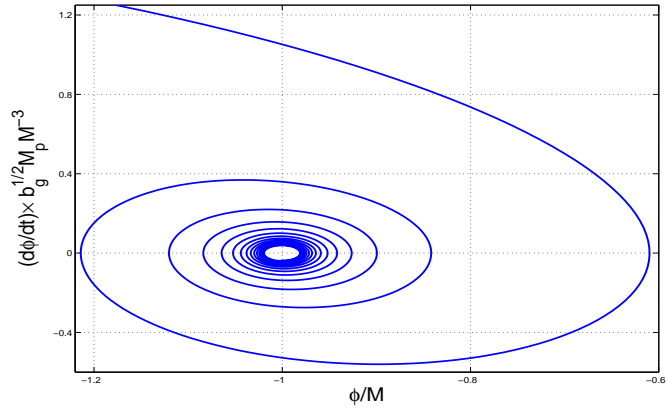}
\end{center}
 \caption{Typical phase curve (in the phase plane ($\phi$,$\dot{\phi})$)
  of the scalar field $\phi$ during
the transition to $\Lambda =0$ state. For illustrative purposes
the parameters are chosen such that
$V_{eff}=(M^2/2b_g)(\phi^2-M^2)^2/(\phi^2+4M^2)$ and $\phi_0=\pm
M$ and $\delta =0$. In the case without fine tuning of the
parameters $b_g\neq b_{\phi}$, i.e. $\delta\neq 0$, the phase
portrait is qualitatively the same.}\label{fig1}
\end{figure}

It follows from the constraint (\ref{constraint2-1}) (where  we
took $\delta =0$) that $|\zeta|\to\infty$ as $\phi\to\phi_0$. More
exactly, oscillations of $sM^{4}+V_1$ around zero are accompanied
with a singular behavior of $\zeta$ each time when $\phi$ crosses
$\phi_0$
\begin{equation}
\frac{1}{\zeta}\sim sM^{4}+V_1(\phi)\to 0 \qquad as \qquad
\phi\to\phi_0 \label{modul-zeta-infty}
\end{equation}
and $\zeta^{-1}$ oscillates around zero together with
$sM^{4}+V_1(\phi)$. Taking into account that the metric in the
Einstein frame $\tilde{g}_{\mu\nu}$, Eq.(\ref{FRW}), is regular we
deduce from Eq.(\ref{gmunuEin}) that the metric $g_{\mu\nu}$ used
in the underlying action (\ref{S-model-scalar.f.}) becomes
degenerate each time when $\phi$ crosses $\phi_0$
\begin{equation}
g_{00}=\frac{\tilde{g}_{00}}{\zeta +b_g}\sim \frac{1}{\zeta}\to 0;
\qquad g_{ii}=\frac{\tilde{g}_{ii}}{\zeta +b_g}\sim
-\frac{1}{\zeta}\to 0 \qquad as \qquad \phi\to\phi_0,
 \label{g00-degen}
\end{equation}
where we have taken into account that the energy density
approaches zero and therefore for this cosmological solution the
scale factor $a(t)$ remains finite in all times $t$. Therefore
\begin{equation}
\sqrt{-g}\sim \frac{1}{\zeta^2}\to 0 \qquad and \qquad \Phi
=\zeta\sqrt{-g}\sim \frac{1}{\zeta}\to 0\qquad as \qquad
\phi\to\phi_0\label{sqrtg-degen}
\end{equation}

 The detailed behavior of $\zeta$, the manifold
measure $\Phi$ and $g_{\mu\nu}$ - components\footnote{Since the
metric in the Einstein frame $\tilde{g}_{\mu\nu}$ is diagonal,
Eq.(\ref{FRW}), it is clear from the transformation
(\ref{gmunuEin}) that $g_{\mu\nu}$ is also diagonal.}  are shown
in Fig. 2.
\begin{figure}[htb]
\includegraphics[width=10.0cm,height=7.0cm]{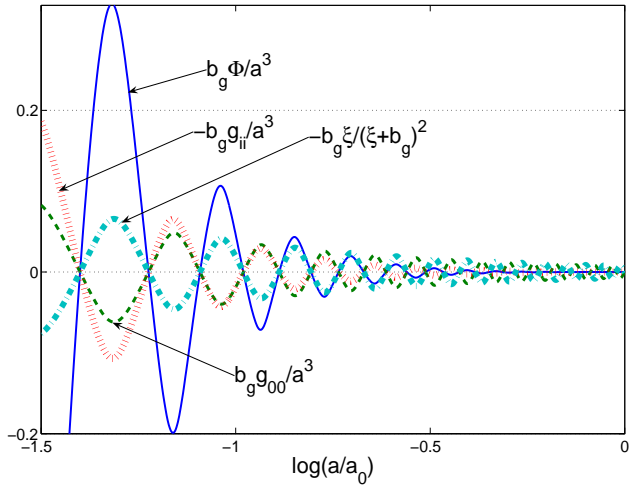}
\caption{Oscillations of the measure $\Phi$, the original metric
$g_{\mu\nu}$ and the r.h.s. of Eq.(\ref{covar-conserv-J}) during
the transition to $\Lambda =0$ state.  }\label{fig2}\end{figure}

Recall that the manifold volume measure $\Phi$ is a signed
measure\cite{signed} and therefore it  is not a surprise that it
can change sign. But TMT shows that including the manifold degrees
of freedom into the dynamics of the scalar-gravity system we
discover an interesting dynamical effect: a transition to zero
vacuum energy is accompanied by oscillations of $\Phi$ around
zero. Similar oscillations\footnote{Note however that these
oscillations do not effect the sign of the metrical volume measure
$g=det(g_{\mu\nu})$ used in the underlying action
(\ref{S-model-scalar.f.}). This notion is useful when comparing
our model with an approach developed in
Refs.\cite{Erdem}-\cite{reflection2}} simultaneously occur with
all components of the metric $g_{\mu\nu}$ used in the underlying
action (\ref{S-model-scalar.f.}).

 The measure $\Phi$ and
the metric $g_{\mu\nu}$ pass zero  only in a discrete set of
moments in the course of  transition to the $\Lambda =0$ state.
Therefore there is no problem with the condition $\Phi\neq 0$ used
for the solution (\ref{varphi}). Also there is no problem with
singularity of $g^{\mu\nu}$ in the underlying action  since
\begin{equation}
\lim_{\phi\to\phi_0}\Phi g^{\mu\nu}=finite \qquad and \qquad
\sqrt{-g} g^{\mu\nu}\sim\frac{1}{\zeta}\to 0 \qquad as \qquad
\phi\to\phi_0 \label{lim}
\end{equation}
The metric in the Einstein frame $\tilde{g}_{\mu\nu}$ is always
regular because degeneracy of $g_{\mu\nu}$ is compensated in
Eq.(\ref{gmunuEin}) by singularity of the ratio of two measures
$\zeta\equiv\Phi/\sqrt{-g}$.

\section{Restoration of Intrinsic TMT Symmetry in the Course of
 Transition to Zero Cosmological Constant State}

Let us now turn to intrinsic symmetry of TMT which can reveal
itself in a model with only the manifold volume measure $\Phi$.
Indeed, if in Eq.(\ref{L1L2}) we take the limit\footnote{for the
model of Sec.IV it means that in Eq.(\ref{S-model-scalar.f.}) we
take the limit $b_g\to 0$, $b_{\phi}\to 0$ and $V_2\to 0$)}
$b_g\to 0$ and $L_2^m\to 0$ then Eq.(\ref{Constr-original}) reads
\begin{equation}
L_1^m-g^{\mu\nu}\frac{\partial L_1^m}{\partial g^{\mu\nu}}=sM^4,
\quad {\text if} \quad\zeta\neq 0. \label{Constr-L2=0}
\end{equation}
If in addition $L_1^m$ is homogeneous of degree 1 in $g^{\mu\nu}$
then the integration constant $M$ must be zero. The simplest
example of a model for $L_1^m$ satisfying this property is the
massless scalar field. In such a case the theory is invariant
under transformations  in the space of the scalar fields
$\varphi_{a}$
\begin{equation}
\varphi_{a}\rightarrow\varphi^{\prime}_{a}=
\varphi^{\prime}_{a}(\varphi_{b}) \label{LES1}
\end{equation}
resulting in the transformation of the manifold volume measure
$\Phi$
\begin{equation}
\Phi(x)\rightarrow\Phi^{\prime}(x)=J(x)\Phi(x), \qquad J(x)=
Det(\frac{\partial\varphi^{\prime}_{a}}{\partial\varphi_{b}})
\label{LES2}
\end{equation}
simultaneously with the local transformation of the metric
\begin{equation}
g_{\mu\nu}(x)\rightarrow g^{\prime}_{\mu\nu}(x)=J(x)g_{\mu\nu}(x).
\label{LES3}
\end{equation}
 This symmetry was studied in earlier
pulications\cite{GK1} where we called it the local Einstein
symmetry (LES).

Consider now linear transformations in the space of the scalar
fields $\varphi_{a}$
\begin{equation}
\varphi_{a}\rightarrow\varphi^{\prime}_{a}= A_a^b\varphi_{b}+C_b,
\quad a,b=1,2,3,4 \label{linear-trans}
\end{equation}
where $A_a^b=constants$, $C_b=constants$. Then LES
(\ref{LES1})-(\ref{LES3}) is reduced to transformations of the
global Einstein symmetry (GES) with $J=det(A_a^b)=const$. Notice
that the Einstein symmetry contains a $\Bbb Z_2$ subgroup of the
sign inversions when $J=-1$:
\begin{equation}
\Phi\rightarrow -\Phi, \quad g_{\mu\nu}\rightarrow
-g_{\mu\nu}\label{reflection}
\end{equation}

LES as well as GES appear to be explicitly broken if $L_1^m$ is
not a homogeneous function of degree 1 in $g^{\mu\nu}$, for
example as in the model where $L_1^m$ describes a scalar field
with a nontrivial potential\footnote{Note that the pure gravity
model of Sec.III is invariant both under the LES and the GES.}.
The Lagrangian $L_2^m$ generically breaks the Einstein symmetry
too. The transformation of GES originated by the infinitesimal
linear transformations
$\varphi_a(x)\rightarrow\varphi^{\prime}_a(x)=
(1+\epsilon/4)\varphi_a(x)$, $\epsilon =const.$, yields the
following variation of the action $(\ref{sec-2-S-modif})$ written
in the form $S=\int{\cal L}d^4x$ where ${\cal L}=\Phi
L_1+\sqrt{-g}L_2$:
\begin{equation}
\delta S=\int\left[-\frac{\partial{\cal L}}{\partial
g^{\mu\nu}}g^{\mu\nu}+
L_1\frac{\partial\Phi}{\partial\varphi_{a,\mu}}\varphi_{a,\mu}\right]\epsilon
d^4x. \label{deltaS}
\end{equation}
The first term in (\ref{deltaS}) equals zero on the mass shell
giving the gravitational equation (\ref{g-mu-nu-varying}); recall
that we proceed in the first order formalism. Integrating the
second term by part, using Eq.(\ref{varphi})  and the definition
(\ref{Phi}) of the measure $\Phi$, we reduce the variation
(\ref{deltaS}) to $\delta S=\epsilon\int\partial_{\mu}j^{\mu}d^4x$
where $\partial_{\mu}j^{\mu}=sM^4\Phi$ and
$j^{\mu}=sM^4B^{\mu}_a\varphi_a$. In the presence of topological
defects with $\Phi =0$, Eq.(\ref{varphi}) does not hold anymore
all over space-time, and one should keep $L_1$ in the definition
of the current: $j^{\mu}=L_1B^{\mu}_a\varphi_a$. In Subs.VIII.D we
 will see how such a situation may be realized.

To present the current conservation in the generally coordinate
invariant form one has to use the covariant divergence. However
when doing this using the original metric $g_{\mu\nu}$ we
encounter the non-metricity. It is much more transparent to use
the Einstein frame (\ref{gmunuEin}) where the space-time becomes
pseudo-Riemannian and the covariant derivative of the metric
$\tilde{g}_{\mu\nu}$ equals zero identically. Thus with the
definition $j^{\mu}=\sqrt{-\tilde{g}}J^{\mu}$, using the
definition of $\zeta$ in Eq.(\ref{g-mu-nu-varying}) and the
transformation to the Einstein frame (\ref{gmunuEin}) we obtain
\begin{equation}
\tilde{\nabla}_{\mu}J^{\mu}\equiv\frac{1}{\sqrt{-\tilde{g}}}\partial_{\mu}
\left(\sqrt{-\tilde{g}}J^{\mu}\right)=sM^4\frac{\zeta}{(\zeta
+b_g)^2} \label{covar-conserv-J}
\end{equation}

As one should expect, when $L_2\equiv 0$ and $L_1^m$ is
homogeneous of degree 1 in $g^{\mu\nu}$, i.e. in the case of
unbroken GES, the current is conserved because in this case the
integration constant $M=0$.

As we have seen in the framework of the scalar field model of
Sec.V, the dynamical evolution  pushes
$|\zeta|\equiv|\Phi|/\sqrt{-g}\to\infty$ as the gravity$+$scalar
field $\phi$ -system approaches (without fine tuning) the $\Lambda
=0$ ground state $\phi =\phi_0$.  Therefore according to
Eq.(\ref{covar-conserv-J}),
\begin{equation}
\tilde{\nabla}_{\mu}J^{\mu}\to 0 \qquad as \qquad \phi\to\phi_0.
\label{covar-conserv-J-vacuum}
\end{equation}
 For the model of Sec.V, the damping oscillations of the r.h.s. of
Eq.(\ref{covar-conserv-J}) around zero are shown in Fig. 2. Thus,
the
 {\em GES explicitly broken in the underlying action, emerges in
the vacuum which, as it turns out, has zero energy density. And
vice versa, emergence of GES due to $|\zeta|\to\infty$ implies,
according to Eq.(\ref{Veff1}), a transition to a $\Lambda =0$
ground state}.

Other way to reach the same conclusion is to look at the
underlying action (\ref{S-model-scalar.f.}). In virtue of
Eq.(\ref{lim}), it is evident that in the course of transition to
the ground state, the terms in (\ref{S-model-scalar.f.}) coupled
to the metric volume measure $\sqrt{-g}$  become negligible in
comparison with the corresponding terms coupled to the manifold
volume measure $\Phi$; besides the  term $-\int V_1(\phi)\Phi
d^4x$ (which also breaks the GES) disappears as $\phi\to\phi_0$.
The only terms surviving in the transition to the $\Lambda =0$
ground state are the following
\begin{equation}
\frac{1}{b_g}\int\Phi d^4x \left[-\frac{1}{\kappa}
R(\Gamma,g)+\frac{1}{2}g^{\mu\nu}\phi_{,\mu}\phi_{,\nu} \right]
\label{S-model-scalar.f.GES}
\end{equation}
and they respect the GES.

One should notice however that one can regard the GES as the
symmetry responsible for a zero CC only if TMT is taken in the
strict framework formulated in Sec.I. In fact, let us consider for
example a modified TMT model where the manifold volume measure
degrees of freedom enter in the Lagrangian $L_1$ in contrast to
our additional basic assumption made  Sec.I ( after
Eq.(\ref{sec-2-S-modif})). Namely let us assume that the
Lagrangian $L_1$  in Eq.(\ref{sec-2-S-modif}) involves a term
proportional to $\Phi/\sqrt{-g}$ that explicitly breaks the
infinite dimensional symmetry (\ref{IDS}). To be more concrete we
consider a model with the action
\begin{equation}
S=S_{mod}^{(1)}-\lambda\int\frac{\Phi^2}{\sqrt{-g}}d^4x
\label{S+Delta}
\end{equation}
where $S_{mod}^{(1)}$ is  the action defined in
Eq.(\ref{S-model-scalar.f.}). Such an addition to the action
(\ref{S-model-scalar.f.}) respects the GES but it is easy to see
that it affects the theory in such a way that without fine tuning
it is impossible generically to reach a zero CC (see Appendix B).

\section{Scalar Field Model II. Global Scale Invariance}

Let us now turn to the analyze of the results of the TMT model
possessing a global scale invariance studied early in
detail\cite{G1}-\cite{GKatz},\cite{GK5}-\cite{GK9}. The scalar
field $\phi$ playing the role of a model of dark energy appears
here as a dilaton, and a spontaneous breakdown of the scale
symmetry results directly from the presence of the manifold volume
measure $\Phi$. In other words, this SSB is an intrinsic feature
of TMT.

In the context of the present paper this model is of significant
interest because  cosmological solutions  of the FRW universe
exhibit  two unexpected results: (a) the ground state as well as
the asymptotic of quintessence-like evolution (in co-moving frame)
possess  certain degeneracies in  $\Phi$ or $g_{\mu\nu}$; (b)
superaccelerating expansion of the universe (phantom cosmology)
appears as the direct dynamical effect  when $\Phi<0$, i.e. as the
orientation of the space-time manifold is opposite to the regular
one. In this section we present the model and some of its relevant
results. Regimes (a) and (b) will be analyzed in the next two
sections.

The action of the model reads
\begin{eqnarray}
&S=&\frac{1}{b_g}\int d^{4}x e^{\alpha\phi /M_{p}}
\left[-\frac{1}{\kappa}R(\Gamma ,g)(\Phi +b_{g}\sqrt{-g})+(\Phi
+b_{\phi}\sqrt{-g})\frac{1}{2}g^{\mu\nu}\phi_{,\mu}\phi_{,\nu}-e^{\alpha\phi
/M_{p}}\left(\Phi V_{1} +\sqrt{-g}V_{2}\right)\right]
 \label{totaction}
\end{eqnarray}
 and it is invariant under the global scale transformations ($\theta =const.$):
\begin{equation}
    g_{\mu\nu}\rightarrow e^{\theta }g_{\mu\nu}, \quad
\Gamma^{\mu}_{\alpha\beta}\rightarrow \Gamma^{\mu}_{\alpha\beta},
\quad \varphi_{a}\rightarrow \lambda_{ab}\varphi_{b}\quad
\text{where} \quad \det(\lambda_{ab})=e^{2\theta}, \quad
\phi\rightarrow \phi-\frac{M_{p}}{\alpha}\theta . \label{st}
\end{equation}
The appearance of the dimensionless parameters $b_g$ and
$b_{\phi}$ is explained by the same reasons we mentioned after
Eqs.(\ref{L1L2}) and (\ref{S-model-scalar.f.}). In contrast to the
model of Sec.IV, now we deal with exponential (pre-) potentials
where $V_1$ and $V_2$ are constant dimensionfull parameters. The
remarkable feature of this TMT model is that Eq.(\ref{varphi}),
being the solution of the equation of motion resulting from
variation of the manifold volume measure degrees of freedom,
breaks spontaneously the scale symmetry (\ref{st}): this happens
due to the appearance of a dimensionfull integration constant
$sM^4$ in Eq.(\ref{varphi}). One can show\cite{GK9} that in the
case of the negative integration constant ($s=-1$) and $V_1>0$,
the ground state appears to be again (as it was in the scalar
field model I of Sec.IV) a zero CC state without fine tuning of
the parameters and initial conditions. The behavior of $\Phi$ and
$g_{\mu\nu}$ in the course of transition to the $\Lambda =0$ state
is qualitatively the same as we observed in Sec.V for the scalar
field model I. Therefore in the present paper studying the model
(\ref{totaction}) we restrict ourself with the choice $s=+1$ and
$V_1>0$.

Similar to the model of Sec.IV, equations of motion resulting from
the action (\ref{totaction}) are noncanonical and the space-time
is non Riemannian when using the original set of variables. This
is because  all the equations of motion and the solution for the
connection coefficients include terms proportional to
$\partial_{\mu}\zeta$. However, when working with the new metric
($\phi$
 remains the same)
\begin{equation}
\tilde{g}_{\mu\nu}=e^{\alpha\phi/M_{p}}(\zeta +b_{g})g_{\mu\nu},
\label{ct}
\end{equation}
which we call the Einstein frame,
 the connection  becomes Riemannian and  general form of all equations
 becomes canonical. Since
$\tilde{g}_{\mu\nu}$ is invariant under the scale transformations
(\ref{st}), spontaneous breaking of the scale symmetry  is reduced
in the Einstein frame to the {\it spontaneous breakdown of the
shift symmetry}
\begin{equation}
 \phi\rightarrow\phi +const.
 \label{phiconst}
\end{equation}

 After the change of
variables  to the Einstein frame (\ref{ct}) the gravitational
equation takes the standard GR form with the same Newton constant
as in the action (\ref{totaction})
\begin{equation}
G_{\mu\nu}(\tilde{g}_{\alpha\beta})=\frac{\kappa}{2}T_{\mu\nu}^{eff}
 \label{gef1}
\end{equation}
where  $G_{\mu\nu}(\tilde{g}_{\alpha\beta})$ is the Einstein
tensor in the Riemannian space-time with the metric
$\tilde{g}_{\mu\nu}$. The energy-momentum tensor
$T_{\mu\nu}^{eff}$ reads
\begin{eqnarray}
T_{\mu\nu}^{eff}&=&\frac{\zeta +b_{\phi}}{\zeta +b_{g}}
\left(\phi_{,\mu}\phi_{,\nu}-\frac{1}{2}
\tilde{g}_{\mu\nu}\tilde{g}^{\alpha\beta}\phi_{,\alpha}\phi_{,\beta}\right)
-\tilde{g}_{\mu\nu}\frac{b_{g}-b_{\phi}}{2(\zeta +b_{g})}
\tilde{g}^{\alpha\beta}\phi_{,\alpha}\phi_{,\beta}
+\tilde{g}_{\mu\nu}V_{eff}(\phi,\zeta;M)
 \label{Tmn}
\end{eqnarray}
where the function $V_{eff}(\phi,\zeta;M)$ is defined as
following:
\begin{equation}
V_{eff}(\phi,\zeta;M)=
\frac{b_{g}\left[M^{4}e^{-2\alpha\phi/M_{p}}+V_{1}\right]
-V_{2}}{(\zeta +b_{g})^{2}}. \label{Veff2}
\end{equation}
 Note that the  $\zeta$-dependence
of $V_{eff}(\phi,\zeta;M)$ is the same as in Eq.(\ref{Veff1}) of
the model of Sec.IV.

The scalar field $\zeta$  is determined by means of the constraint
similar to Eq.(\ref{constraint2-1}) of Sec.IV
\begin{eqnarray}
&&(b_{g}-\zeta)\left[M^{4}e^{-2\alpha\phi/M_{p}}+
V_{1}\right]-2V_{2}-\delta\cdot b_{g}(\zeta +b_{g})X
=0\label{constraint2}
\end{eqnarray}
where
\begin{equation}
X\equiv\frac{1}{2}\tilde{g}^{\alpha\beta}\phi_{,\alpha}\phi_{,\beta}
\qquad \text{and} \qquad \delta =\frac{b_{g}-b_{\phi}}{b_{g}}
\label{delta}
\end{equation}

The dilaton $\phi$ field equation in the Einstein frame is reduced
to the following
\begin{equation}
\frac{1}{\sqrt{-\tilde{g}}}\partial_{\mu}\left[\frac{\zeta
+b_{\phi}}{\zeta
+b_{g}}\sqrt{-\tilde{g}}\tilde{g}^{\mu\nu}\partial_{\nu}\phi\right]-\frac{2\alpha\zeta}{(\zeta
+b_{g})^{2}M_{p}}M^{4}e^{-2\alpha\phi/M_{p}} =0.
\label{phi-after-con}
\end{equation}
where again $\zeta$  is a solution of the constraint
(\ref{constraint2}). Note that the dilaton $\phi$ dependence in
all equations of motion in the Einstein frame appears {\em only}
in the form $M^{4}e^{-2\alpha\phi/M_{p}}$, i.e. it results only
from the spontaneous breakdown of the scale symmetry (\ref{st}).

The effective energy-momentum tensor (\ref{Tmn}) can be
represented in a form of that of  a perfect fluid
\begin{equation}
T_{\mu\nu}^{eff}=(\rho +p)u_{\mu}u_{\nu}-p\tilde{g}_{\mu\nu},
\qquad \text{where} \qquad
u_{\mu}=\frac{\phi_{,\mu}}{(2X)^{1/2}}\label{Tmnfluid}
\end{equation}
with the following energy and pressure densities resulting from
Eqs.(\ref{Tmn}) and (\ref{Veff2}) after inserting the solution
$\zeta =\zeta(\phi,X;M)$ of Eq.(\ref{constraint2}):
\begin{equation}
\rho(\phi,X;M) =X+ \frac{(M^{4}e^{-2\alpha\phi/M_{p}}+V_{1})^{2}-
2\delta b_{g}(M^{4}e^{-2\alpha\phi/M_{p}}+V_{1})X -3\delta^{2}
b_{g}^{2}X^2}{4[b_{g}(M^{4}e^{-2\alpha\phi/M_{p}}+V_{1})-V_{2}]},
\label{rho1}
\end{equation}
\begin{equation}
p(\phi,X;M) =X- \frac{\left(M^{4}e^{-2\alpha\phi/M_{p}}+V_{1}+
\delta b_{g}X\right)^2}
{4[b_{g}(M^{4}e^{-2\alpha\phi/M_{p}}+V_{1})-V_{2}]}. \label{p1}
\end{equation}

In a spatially flat FRW universe with the metric
$\tilde{g}_{\mu\nu}=diag(1,-a^2,-a^2,-a^2)$ filled with the
homogeneous scalar field $\phi(t)$, the $\phi$  field equation of
motion takes the form
\begin{equation}
Q_{1}\ddot{\phi}+ 3HQ_{2}\dot{\phi}- \frac{\alpha}{M_{p}}Q_{3}
M^{4}e^{-2\alpha\phi/M_{p}}=0 \label{phi1}
\end{equation}
 where $H$ is the Hubble parameter and we have used the following notations
\begin{equation}
\dot{\phi}\equiv \frac{d\phi}{dt} \label{phidot-vdot}
\end{equation}
\begin{equation}
Q_1=2[b_{g}(M^{4}e^{-2\alpha\phi/M_{p}}+V_{1})-V_{2}]\rho_{,X}
=(b_{g}+b_{\phi})(M^{4}e^{-2\alpha\phi/M_{p}}+V_{1})-
2V_{2}-3\delta^{2}b_{g}^{2}X \label{Q1}
\end{equation}
\begin{equation}
Q_2=2[b_{g}(M^{4}e^{-2\alpha\phi/M_{p}}+V_{1})-V_{2}]p_{,X}=
(b_{g}+b_{\phi})(M^{4}e^{-2\alpha\phi/M_{p}}+V_{1})-
2V_{2}-\delta^{2}b_{g}^{2}X\label{Q2}
\end{equation}
\begin{equation}
Q_{3}=\frac{1}{[b_{g}(M^{4}e^{-2\alpha\phi/M_{p}}+V_{1})-V_{2}]}
\left[(M^{4}e^{-2\alpha\phi/M_{p}}+V_{1})
[b_{g}(M^{4}e^{-2\alpha\phi/M_{p}}+V_{1})-2V_{2}] +2\delta
b_{g}V_{2}X+3\delta^{2}b_{g}^{3}X^{2}\right] \label{Q3}
\end{equation}

The non-linear $X$-dependence appears here in the framework of the
fundamental theory without exotic terms in the Lagrangians $L_1$
and $L_2$.  This effect follows just from the fact that there are
no reasons to choose the parameters $b_{g}$ and $b_{\phi}$ in the
action (\ref{totaction}) to be equal in general; on the contrary,
the choice $b_{g}=b_{\phi}$ would be a fine tuning.  Thus the
above equations represent an  explicit example of
$k$-essence\cite{k-essence}  resulting from first principles. The
system of equations (\ref{gef}), (\ref{rho1})-(\ref{phi1})
accompanied with the functions (\ref{Q1})-(\ref{Q3}) and written
in the metric $\tilde{g}_{\mu\nu}=diag(1,-a^2,-a^2,-a^2)$ can be
obtained from the k-essence type effective action
\begin{equation}
S_{eff}=\int\sqrt{-\tilde{g}}d^{4}x\left[-\frac{1}{\kappa}R(\tilde{g})
+p\left(\phi,X;M\right)\right] \label{k-eff},
\end{equation}
where $p(\phi,X;M)$ is given by Eq.(\ref{p1}). In contrast to the
simplified models studied in literature\cite{k-essence}, it is
impossible here to represent $p\left(\phi,X;M\right)$ in a
factorizable form like $\tilde{K}(\phi)\tilde{p}(X)$. The scalar
field effective Lagrangian, Eq.(\ref{p1}), can be represented in
the form
\begin{equation}
p\left(\phi,X;M\right)=K(\phi)X+ L(\phi)X^2-U(\phi)
\label{eff-L-ala-Mukhanov}
\end{equation}
where the potential
\begin{equation}
U(\phi)=\frac{[V_{1}+M^{4}e^{-2\alpha\phi/M_{p}}]^{2}}
{4[b_{g}\left(V_{1}+M^{4}e^{-2\alpha\phi/M_{p}}\right)-V_{2}]}
 \label{eff-L-ala-Mukhanov-potential}
\end{equation}
and $K(\phi)$ and $L(\phi)$ depend on $\phi$ only via
$M^{4}e^{-2\alpha\phi/M_{p}}$. Notice that $U(\phi)>0$  for any
$\phi$ provided
\begin{equation}
b_g>0, \qquad V_1>0 \qquad and \qquad b_{g}V_{1}\geq V_{2}
\label{bV1>V2},
\end{equation}
that we will assume in what follows.  Note  that besides the
presence of the effective potential $U(\phi)$, the Lagrangian
$p\left(\phi,X;M\right)$ differs from that of
Ref.\cite{k-inflation-Mukhanov} by the sign of $L(\phi)$: in our
case $L(\phi)<0$ provided the conditions (\ref{bV1>V2}). This
result cannot be removed by a choice of the parameters of the
underlying action (\ref{totaction}) while in
Ref.\cite{k-inflation-Mukhanov} the positivity of $L(\phi)$ was an
essential {\it assumption}. This difference plays a crucial role
for a possibility of a dynamical protection from the initial
singularity of the curvature studied in detail in Ref\cite{GK9}.

The model allows a power law inflation (where the dilaton $\phi$
plays the role of the inflaton) with a graceful exit to a zero or
tiny cosmological constant state.  In what it concerns to
primordial perturbations of $\phi$ and their evolution, there are
no difference with the usual (i.e. one-measure) model with the
action (\ref{k-eff})-(\ref{eff-L-ala-Mukhanov-potential}).

In the model under consideration, the conservation law
corresponding to the GES (\ref{linear-trans}) has the form
\begin{equation}
\tilde{\nabla}_{\mu}J^{\mu}\equiv\frac{1}{\sqrt{-\tilde{g}}}\partial_{\mu}
\left(\sqrt{-\tilde{g}}J^{\mu}\right)=sM^4\frac{\zeta}{(\zeta
+b_g)^2}e^{-2\alpha\phi/M_{p}} \label{covar-conserv-J-model2}
\end{equation}
with the same definition of the current $J^{\mu}$ as in Sec.VI.

\section{Degeneracies of $g_{00}$ and $\Phi$ in
$\Lambda\neq 0$ Ground States }

\subsection{Fine-Tuned $\delta =0$ Models}

We are going now to analyze some of the cosmological solutions for
the late universe in the framework of the scale invariant model of
the previous section. These solutions surprisingly exhibit that
asymptotically, as $t\to\infty$, either $g_{00}\to 0$ or $\Phi\to
0$.

In the late universe, the kinetic energy $X\to 0$. Therefore in
many cases the role of the nonlinear $X$ dependence becomes
qualitatively unessential. This is why, for simplicity, in this
section we can restrict ourself with the fine tuned model with
$\delta =0$. In such a case the constraint (\ref{constraint2})
yields
\begin{equation}
\zeta
=\frac{b_gV_1-2V_2+b_gM^{4}e^{-2\alpha\phi/M_{p}}}{V_1+M^{4}e^{-2\alpha\phi/M_{p}}},
\label{zeta-without-ferm-delta=0}
 \end{equation}
 The energy density and pressure take then the form
\begin{equation}
\rho^{(0)}=\rho|_{\delta =0}=\frac{1}{2}\dot{\phi}^{2}+U(\phi);
\qquad p^{(0)}=p|_{\delta =0}=\frac{1}{2}\dot{\phi}^{2}-U(\phi),
\label{rho-delta=0}
\end{equation}
where $U(\phi)$ is determined by
Eq.(\ref{eff-L-ala-Mukhanov-potential}). The $\phi$-equation
(\ref{phi1}) is reduced to
\begin{equation}
\ddot{\phi}+3H\dot{\phi}+\frac{dU(\phi)}{d\phi}=0.
\label{eq-phief-without-ferm-delta=0}
\end{equation}

 Applying
this model to  of the late time cosmology
 of the spatially flat universe and assuming that the scalar field
$\phi\rightarrow\infty$ as $t\rightarrow\infty$, it is convenient
to rewrite the potential $U(\phi)$ in the form
\begin{equation}
U(\phi)=\Lambda +V(\phi), \label{rho-without-ferm}
\end{equation}
where
\begin{equation}
 \Lambda
=\frac{V_{1}^{2}} {4(b_{g}V_{1}-V_{2})}. \label{lambda}
\end{equation}
 is the positive cosmological constant and
\begin{equation}
V(\phi)
=\frac{(b_{g}V_{1}-2V_{2})V_{1}M^{4}e^{-2\alpha\phi/M_{p}}+
(b_{g}V_{1}-V_{2})M^{8}e^{-4\alpha\phi/M_{p}}}
{4(b_{g}V_{1}-V_{2})[b_{g}(V_{1}+
M^{4}e^{-2\alpha\phi/M_{p}})-V_{2}]}.
\label{V-quint-without-ferm-delta=0}
\end{equation}
 It is evident that if $b_{g}V_1>
2V_2$ or $b_{g}V_1= 2V_2$ then $V(\phi)$ is a sort of a
quintessence-like potential and therefore quintessence-like
scenarios can be realized. This means that the dynamics of the
late time universe is governed by the dark energy which consists
of both the cosmological constant and the potential slow decaying
to zero as $\phi\to\infty$. In the opposite case, $b_{g}V_1<2V_2$,
the potential $V(\phi)$, and also $U(\phi)$, has an absolute
minimum at some finite value of $\phi$, and therefore the
cosmological scenario is different from the quintessence-like.
Details of the cosmological evolution starting from the early
inflation and up to the late time universe governed by the
potential $U(\phi)$ have been studied in Ref.\cite{GK9} for each
of these three cases. Here we want to analyze what kind of
degeneracy appears  in ground state depending on the region in the
parameter space.

\subsection{The case $b_{g}V_1>2V_2$}

Let us consider the case when the relation between the parameters
$V_1$ and $V_2$ satisfies the condition $b_{g}V_1>2V_2$.  It
follows from Eq.(\ref{zeta-without-ferm-delta=0}) that
\begin{equation}
\zeta \to\frac{b_gV_1-2V_2}{V_1}=const
> 0 \quad as \quad \phi\to\infty \label{zeta-asympt-delta=0}
 \end{equation}

By making use the  $(00)$ component of Eq.(\ref{ct}), we see that
\begin{equation}
g_{00}=\frac{e^{-\alpha\phi/M_{p}}}{\zeta +b_g}\to 0
\label{g00-general-asympt-delta=0}
 \end{equation}
 In order to get the asymptotic time dependence of $g_{00}$ and
   the spatial components of the metric
\begin{equation}
 g_{ii}=-\frac{e^{-\alpha\phi/M_{p}}}{\zeta +b_g}\,a(t)^2,
\qquad i=1,2,3
 \label{g-phi-sqrt-general-asympt-delta=0}
 \end{equation}
as $t\to\infty$, we have to know a solution $a=a(t)$, $\phi
=\phi(t)$. We can find analytically the asymptotic (as
$\phi\to\infty$) behavior of a cosmological solution for a
particular value of the parameter $\alpha =\sqrt{3/8}$. In such a
case, keeping only the leading contribution of the $\phi$-exponent
in Eq.(\ref{V-quint-without-ferm-delta=0}), we deal with the
following system of equations
\begin{equation}
\left(\frac{\dot{a}}{a}\right)^{2}=\frac{1}{3M_{p}^{2}}\rho^{(0)}
 \label{FRW-eq1}
\end{equation}
\begin{equation}
\ddot{\phi}+3\frac{\dot{a}}{a}\dot{\phi} -\frac{2\alpha
}{M_{p}}\frac{V_1(b_gV_1-2V_2)}{4(b_gV_1-V_2)^2}M^{4}e^{-2\alpha\phi/M_{p}}
=0, \label{phi-eq-cosm}
\end{equation}
where $\rho^{(0)}$ is determined by Eq.(\ref{rho-delta=0}). The
exact analytic solution for these equations is as
follows\cite{GK3}:
\begin{equation}
\phi(t)=const. + \frac{M_p}{2\alpha}\ln(M_pt), \qquad a(t)\propto
t^{1/3}e^{\lambda t}, \qquad \lambda
=\frac{1}{M_p}\sqrt{\frac{\Lambda}{3}} \label{a-phi-exact}
\end{equation}
where $\Lambda$ is determined by Eq.(\ref{lambda}). Therefore we
obtain for the asymptotic cosmic time behavior of the components
of the metric $g_{\mu\nu}$
\begin{equation}
g_{00}\sim\frac{1}{t^{1/2}}\to 0; \qquad g_{ii}\sim
-t^{1/6}e^{2\lambda t} \qquad as \qquad t\to\infty
\label{g00-gii-exact-asympt-delta=0}
 \end{equation}
So in the course of the expansion of the very late universe, only
$g_{00}$  asymptotically vanishes while the space components
$g_{ii}$ behave qualitatively in the same manner as the
 space components of the metric in the Einstein frame
 $\tilde{g}_{ii}$. Respectively, the asymptotic behavior of the volume
 measures is as follows:
\begin{equation}
\Phi\approx \frac{b_{g}V_1-2V_2}{V_{1}}\sqrt{-g}\sim e^{3\lambda
t} \qquad as \qquad t\to\infty \label{Phi-g-exact-asympt-delta=0}
 \end{equation}

 The GES is asymptotically restored that can be seen from the
 asymptotic time behavior of the conservation law (\ref{covar-conserv-J-model2})
 \begin{equation}
\tilde{\nabla}_{\mu}J^{\mu}\sim const\cdot
e^{-2\alpha\phi/M_{p}}\sim \frac{1}{t}.
\label{covar-conserv-J-model2-1}
\end{equation}

 \subsection{The case $b_{g}V_1=2V_2$}

 In this case the asymptotic form of $V(\phi)$ is
\begin{equation}
V(\phi) \approx\frac{M^8}{2b_gV_1}e^{-4\alpha\phi/M_{p}}
\label{V-bgV1equal2V2-delta=0}
\end{equation}
and  $\zeta$ asymptotically approaches zero according to
\begin{equation}
\zeta
=\frac{b_gM^4e^{-2\alpha\phi/M_p}}{V_1+M^4e^{-2\alpha\phi/M_p}}
\to 0 \qquad as \qquad \phi\to\infty.
\label{zeta-asympt-delta=0-1}
 \end{equation}
 Similar to the previous subsection, the analytic form of the
  asymptotic (as $\phi\gg M_p$) cosmological solution exists for a
 particular value of the parameter $\alpha
 =\sqrt{3/32}$:
\begin{equation}
\phi(t)=const. + \frac{M_p}{4\alpha}\ln(M_pt), \qquad a(t)\propto
t^{1/3}e^{\lambda t}, \qquad \lambda
=\frac{1}{M_p}\sqrt{\frac{\Lambda}{3}},
\label{a-phi-exact-bV1equal2V2}
\end{equation}
where now
\begin{equation}
\Lambda =\frac{V_1}{2b_g} \label{Lambda-bV1equal2V2}
\end{equation}

For this solution we obtain the following asymptotic cosmic time
behavior for the components of the metric $g_{\mu\nu}$ and volume
measures:
\begin{equation}
g_{00}\sim\frac{1}{t^{1/4}}\to 0; \qquad g_{ii}\sim
-t^{5/12}e^{2\lambda t} \qquad as \qquad t\to\infty
\label{bgV1equal2V2g00-gii}
 \end{equation}
 \begin{equation}
\sqrt{-g}\sim \sqrt{t}e^{3\lambda t}, \qquad \Phi\sim e^{3\lambda
t} \qquad as \qquad t\to\infty \label{bgV1equal2V2Phi-g}
\end{equation}

The
 asymptotic time behavior of the conservation law describing
 the asymptotic restoration of the GES is the same as in
 Eq.(\ref{covar-conserv-J-model2-1}).

\subsection{The case $0<b_{g}V_1<2V_2$}

In this case the potential $U(\phi)$,
Eq.(\ref{eff-L-ala-Mukhanov-potential}), has an absolute minimum
\begin{equation}
\Lambda =U(\phi_{min})=\frac{V_{2}}{b_{g}^{2}} \qquad \text{at}
\qquad \phi =\phi_{min}=
-\frac{M_{p}}{2\alpha}\ln\left(\frac{2V_{2}-b_{g}V_{1}}{b_{g}M^4}\right).
\label{minVeff}
\end{equation}
The spatially flat universe described in the Einstein frame with
the metric $\tilde{g}_{\mu\nu}=diag(1,-a^2,-a^2,-a^2)$, in a
finite time\cite{GK2} reaches this ground state where it expands
exponentially
\begin{equation}
a\propto e^{\lambda t}, \qquad \lambda =M_p^{-1}\sqrt{\Lambda/3}
\label{a(t)minVeff}
\end{equation}
and $\Lambda$ is given by Eq.(\ref{minVeff}). A surprising feature
of this case is that $\zeta$,
Eq.(\ref{zeta-without-ferm-delta=0}), disappears in the minimum:
\begin{equation}
\zeta(\phi_{min})=0 \label{zeta-in-minVeff}
\end{equation}

The components of the metric $g_{\mu\nu}$ in the ground state are
as follows
\begin{equation}
g_{00}|_{(ground\,state)}=\left(\frac{2V_2-b_gV_1}{b_g^3M^4}\right)^{1/2}=const,
\qquad g_{ii}|_{(ground\,state)}= -g_{00}|_{(ground\,state)}\cdot
e^{2\lambda t} \label{bgV1less2V2g00-gii}
 \end{equation}
with the  respective behavior of the metrical volume measure
$\sqrt{-g}\propto exp(3\lambda t)$. Hence the manifold volume
measure in the ground state disappears
\begin{equation}
\Phi|_{(ground\,state)}=0 \label{bgV1less2V2Phi}
 \end{equation}
 in view of Eq.(\ref{zeta-in-minVeff}).

 Disappearance of the manifold volume measure $\Phi$ in the
   ground state may not allow to get the equation (\ref{varphi})
   by varying the $\varphi_a$ fields in the action (\ref{totaction}).
 Therefore in the conservation law
 (\ref{covar-conserv-J-model2})  one should use the current in the form
$j^{\mu}=L_1B^{\mu}_a\varphi_a$ as we have noticed after
Eq.(\ref{deltaS}). Recall that $L_1$ is constituted by the terms
of the Lagrangian in (\ref{totaction}) coupled to the measure
$\Phi$. However, after using the gravitational equation obtained
by varying $g^{\mu\nu}$ in (\ref{totaction}) and substituting the
ground state value $\phi =\phi_{min}$ into $L_1$, we obtain
$L_1=M^4$. Hence, the conservation law
 (\ref{covar-conserv-J-model2}) in the ground state reads just
\begin{equation}
\tilde{\nabla}_{\mu}J^{\mu}|_{(ground\,state)}=0.
\label{covar-conserv-J-model2-gs}
\end{equation}

\begin{figure}[htb]
\begin{center}
\includegraphics[width=15.0cm,height=10.0cm]{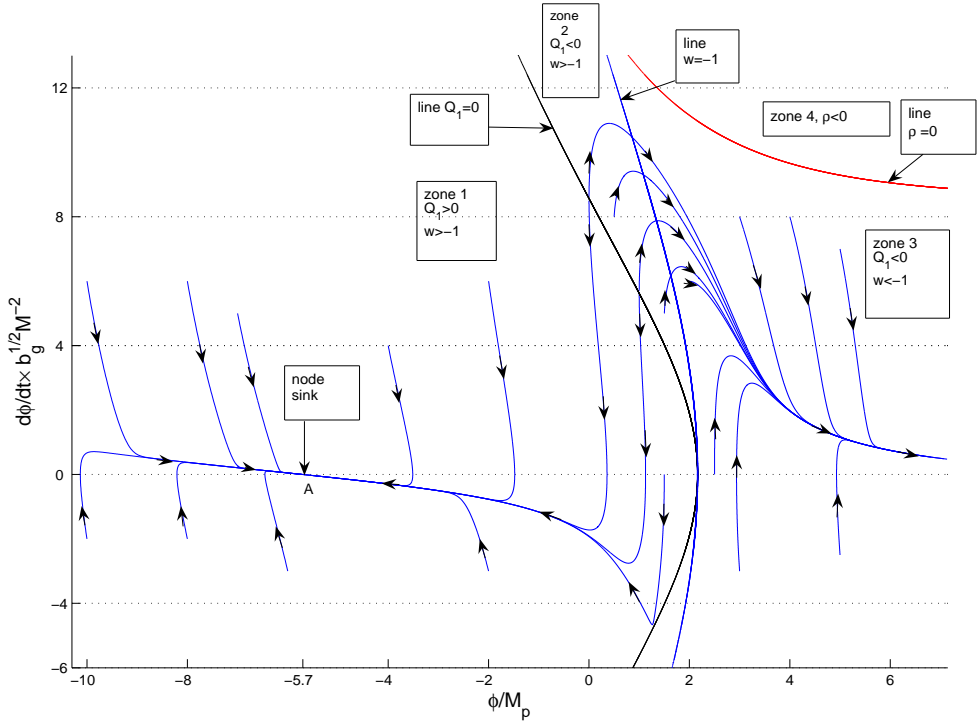}
\end{center}
\caption{The phase portrait (in the phase plane
($\phi$,$\dot{\phi})$) for the model with $\alpha =0.2$, $\delta
=0.1$, $V_{1}=10M^{4}$ and $V_{2}=9.9b_{g}M^{4}$. The region with
$\rho >0$ is divided into two dynamically disconnected regions by
the line $Q_{1}(\phi,\dot{\phi})=0$. To the left of this line
$Q_{1}>0$ (the appropriate zone we call zone 1) and to the right
\, $Q_{1}<0$.  The $\rho
>0$ region to the right of the line $Q_{1}(\phi,\dot{\phi})=0$ is
divided into two zones (zone 2 and zone 3) by the line $Q_2=0$
(the latter coincides with the line where $w=-1$). In zone 2 \,
$w>-1$ but $c_s^2<0$. In zone 3 \,$w<-1$ and $c_s^2>0$. Phase
curves started in zone 2 cross the line $w=-1$. All phase curves
in zone 3 exhibit processes with super-accelerating expansion of
the universe. Besides all the phase curves in zone 3 demonstrate
dynamical attractor behavior to the line which asymptotically, as
$\phi\to\infty$, approaches  the straight line
$\dot{\phi}=0$.}\label{fig3}
\end{figure}
\begin{figure}[htb]
\begin{center}
\includegraphics[width=16.0cm,height=7.0cm]{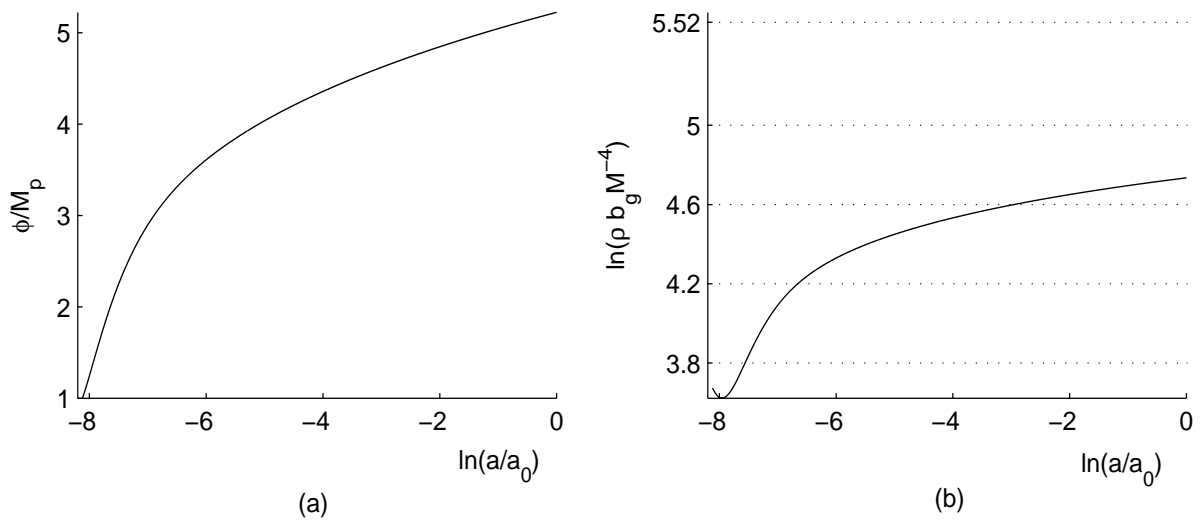}
\end{center}
\caption{For the model with $\alpha =0.2$, $\delta =0.1$,
$V_{1}=10M^{4}$ and $V_{2}=9.9b_{g}M^{4}$: typical scalar factor
dependence of $\phi$  (Fig.(a)) and
 of the energy density $\rho$, defined by Eq.(\ref{rho1}),
(Fig(b)) in the regime corresponding to the phase curves started
in zone 2. Both graphs correspond to the initial conditions
$\phi_{in}=M_{p}$, $\dot\phi_{in} =5.7M^2/\sqrt{b_g}$; $\rho$
increases approaching asymptotically the value
$\frac{M^{4}}{b_{g}}e^{5.52}$}\label{fig4}
\end{figure}
\begin{figure}[htb]
\includegraphics[width=10.0cm,height=8.0cm]{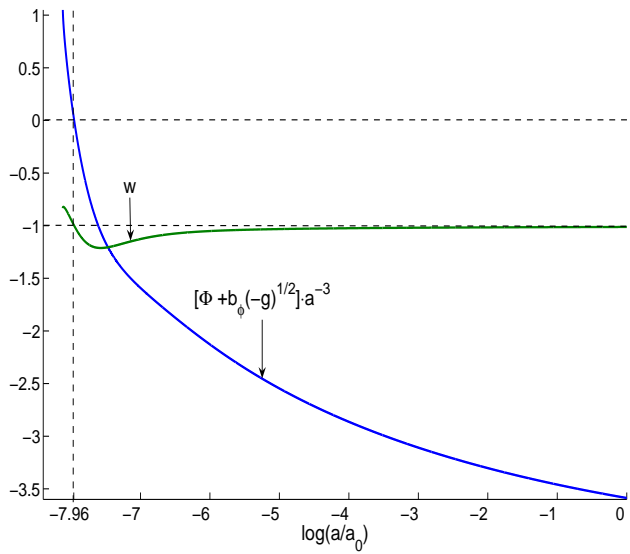}
\caption{For the same model and with the same initial conditions
as in Fig. 4: crossing the phantom divide $w=-1$ and changing sign
of the total volume measure $(\Phi +b_{\phi}\sqrt{-g})$ in the
scalar field $\phi$ kinetic term (in the underlying action
(\ref{totaction})) occur simultaneously.}\label{fig5}
\end{figure}

\section{Sign Indefiniteness of the Manifold Volume Measure
\newline
 as the Origin of a Phantom Dark Energy}

We turn now to the non fine-tuned case of the model of Sec.VII
applied to the spatially flat universe. We start from a short
review of our recent results\cite{GK9} concerning  qualitative
structure of the appropriate dynamical system which consists of
Eq.(\ref{phi1}) and the equation
\begin{equation}
\left(\frac{\dot{a}}{a}\right)^{2}=\frac{1}{3M_{p}^{2}}\rho
 \label{FRW-eq2}
\end{equation}
where the energy density $\rho$ is defined by Eq.(\ref{rho1}). The
case of the interest of this section is realized when the
parameters of the model satisfy the condition
\begin{equation}
(b_g+b_{\phi})V_1-2V_2<0 \label{cond-for-phase-structure-1}
\end{equation}
In this case the phase plane has a very interesting structure
presented in Fig.3. Recall that the functions $Q_1$, $Q_2$, $Q_3$
are defined by Eqs.(\ref{Q1})-(\ref{Q3}).

We are interested in the equation of state $w=p/\rho<-1$, where
 pressure $p$ and energy density $\rho$ are given by
Eqs.(\ref{rho1}) and (\ref{p1}). The line indicated in Fig. 3 as
"line $w=-1$" coincides with the line $Q_2(\phi,X)=0$ because
\begin{equation}
w+1=\frac{X}{\rho}\cdot\frac{Q_2}{\left[b_g\left(M^4e^{-2\alpha\phi/M_p}+V_1\right)-V_2\right]}
\label{w+1}
\end{equation}
 Phase curves in zone 3 correspond to the
cosmological solutions  with the equation of state $w<-1$. In zone
2, $w>-1$ but this zone has no physical meaning since the squared
sound speed of perturbations
\begin{equation}
c_s^2=\frac{Q_2}{Q_1} \label{sound}
\end{equation}
is negative in zone 3. But in zone 2, $c_s^2>0$. Some details of
 numerical solutions describing the cross of the phantom divide
$w=-1$ and the super-accelerating expansion of the universe are
presented in Figs. 4 and 5.

Note that the superaccelerating cosmological expansion is obtained
here without introducing an explicit phantom scalar field into the
underlying action (\ref{totaction}). In Ref.\cite{GK9} we have
discussed this effect from the point of view of the effective
k-essence model realized in the Einstein frame when starting from
the action (\ref{totaction}). A deeper analysis of the same effect
yields the conclusion that the true and profound {\em origin of
the appearance of an effective phantom dynamics in our model is
sign-indefiniteness of the manifold volume measure $\Phi$}. In
fact, using the constraint (\ref{constraint2}), Eqs.(\ref{w+1})
and (\ref{ct}) it is easy to show that
\begin{equation}
\Phi
+b_{\phi}\sqrt{-g}=(w+1)\,\frac{\rho}{4X}\,\frac{[M^4e^{-2\alpha\phi/M_p}+V_1+\delta\cdot
b_gX]}{[b_g(M^4e^{-2\alpha\phi/M_p}+V_1)-V_2]}\, a^3,
\label{kinetic-measure}
\end{equation}
where $a$ is the scale factor. The expression in the l.h.s of this
equation is the total volume measure of the $\phi$ kinetic term in
the underlying action (\ref{totaction}):
\begin{equation}
\int d^{4}x (\Phi
+b_{\phi}\sqrt{-g})\frac{1}{2}g^{\mu\nu}\phi_{,\mu}\phi_{,\nu}
\label{kinetic-total-measure}
\end{equation}
 The sign of this volume measure
coincides with the sign of $w+1$ as well as with the sign of the
function $Q_2$ (see Eq.(\ref{w+1})). In Fig. 5 we present the
result of numerical solution for the scale factor dependence of
$w$ and $(\Phi +b_{\phi}\sqrt{-g})/a^3$.  Thus {\em crossing the
phantom divide occurs when  the total volume measure of the $\phi$
kinetic term in the underlying action changes sign from positive
to negative for dynamical reasons}. This dynamical effect appears
here as a dynamically well-founded alternative to the usually
postulated phantom kinetic term of a scalar field
Lagrangian\cite{Phantom-usual}.

\section{Summary and Discussion}

Introducing the space-time manifold volume element (\ref{dV}) and
adding the appropriate degrees of freedom to a set of traditional
variables (metric, connection, matter fields) we reveal that such
a two measures theory (TMT) takes up a special position between
alternative theories. First, the equations of motion can be
rewritten in the Einstein frame (where the space-time becomes
Riemannian) with {\em the same Newtonian constant} as in the
underlying action (where the space-time is generically
non-Riemannian). Second, the theory possesses remarkable features
in what it concerns the CC problem. Third, the TMT model with
spontaneously broken dilatation symmetry satisfies all existing
tests of GR. There are other interesting results, for example a
possibility of a dynamical protection from the initial singularity
of the curvature.

In this paper we have studied the behavior of the manifold volume
measure $\Phi$ and the metric tensor $g_{\mu\nu}$ (used in the
underlying TMT action) in cosmological solutions for a number of
scalar field models of dark energy. We have made a  special accent
on the {\em sign indefiniteness of the manifold volume measure}
 $\Phi$ that may yield interesting physical effects.
An example of such type of effects we have seen in Sec.IX: the
total volume measure of the dilaton scalar field kinetic term in
the underlying action can change sign from positive to negative in
the course of dynamical evolution of the late time universe. In
the Einstein frame, this transition corresponds to {\em the
crossing of the phantom divide of the dark energy}.

 We have found out that in all
studied models, the transition to the ground state is {\em always
accompanied by a certain degeneracy}  either in the metric (e.g.,
in $g_{00}$ or in all components) or in the manifold volume
measure $\Phi$, or even in both of them. This result differs
sharply from what was expected e.g. in
Refs.\cite{Hawking1979}-\cite{Tseytlin1982} where degenerate
metric solutions have been associated with high curvature and
temperature phases. One should only take into account that
degeneracy of $g_{\mu\nu}$ and/or $\Phi$ in the (transition to)
ground state takes place only when one works with the set of
variables of the underlying TMT action. In the Einstein frame, we
deal with the effective picture where the measure $\Phi$ does not
present at all and the metric tensor $\tilde{g}_{\mu\nu}$ (see
Eqs.(\ref{gmunuEin}) or (\ref{ct}))
 has the same regularity properties as in GR. The regularity of
 $\tilde{g}_{\mu\nu}$ results from the
singularity of the transformations (\ref{gmunuEin}) or
(\ref{ct})): degeneracy of $g_{\mu\nu}$ in a discrete set of
moments is compensated by a singularity of $\zeta$.

\subsection{The CC problem}
 TMT provides two different
 possibilities for resolution of the CC problem: one which guarantees zero CC without fine tuning
 (see however the end of Sec.VI and Appendix B); another which
 allows an unexpected way to reach a tiny CC. Which of these possibilities
 is realized depends on the sign of the integration constant $sM^4$, $s=\pm 1$.
  We are going now to discuss these
 two issues.
\subsubsection{The case  $\Lambda =0$ in TMT}
This case is of a special interest for two reasons. First, as it
was shown earlier\cite{GK9},
 the conditions of the Weinberg's no-go
theorem\cite{Weinberg1} fail and a transition to a zero CC state
in TMT can be realized without fine tuning. This becomes possible
for example if $V_1(\phi)>0$ and the integration constant
$sM^4<0$. Second, as we have shown in Sec.V, in the course of
transition to a zero CC state, $g_{\mu\nu}$ and $\Phi$ oscillate
synchronously around zero and they cross zero  each time $t_i$
($i=1,2,3,...$) when the scalar field $\phi$ crosses the (zero)
absolute minimum of the potential (\ref{Veff3}) (or of the
potential (\ref{eff-L-ala-Mukhanov-potential}) for the model of
Sec.VII with $V_1<0$, see \cite{GK9}).

One should recall that $\zeta(x)$ does not have its own dynamics:
its values at the space-time point $x$ are determined directly and
immediately by the local configuration of the matter fields and
gravity through the algebraic constraint, which is nothing but a
consistency condition of equations of motion. $\zeta(x)$ does not
possess inertia and therefore it changes together and
synchronously with changing matter and gravity fields. This notion
is very important when trying to answer the natural question: can
oscillations of $\zeta(x)$ be a source for particle creation? The
answer is - no, it cannot. In fact, there is a coupling of $\zeta$
with fermions. But the structure of this coupling in the Einstein
frame has very surprising features which we will shortly review in
the next subsection. Here we are only formulating the conclusion:
emergence of even a tiny amount of fermionic matter immediately
yields a rearrangement of the vacuum\footnote{A possibility of a
vacuum deformation in a different approach has been shown by
MacKenzie, Wilczek and Zee\cite{Zee}} in such a way that $\zeta$
instantly ceases the regime of oscillations and rapidly enters
into a regime of monotonous approach to a nonzero constant. It is
interesting to note that the latter effect may explain why the
present day cosmological constant most likely is tiny but nonzero,
in spite of the existence of  a fine tuning free classical
solution described a transition to the $\Lambda =0$ state.

 An overall change of sign of $g_{\mu\nu}$ in the
course of these oscillations means  a change of the signature from
$(+---)$ to $(-+++)$ and vice versa, while oscillations  of the
sign of $\Phi$ describe the change of orientation of the
space-time manifold. The latter means that the arena of the
gravitational dynamics should contain two space-time manifolds
with opposite orientations. The discrete set of changes of the
orientations happens in the form of a {\em smooth} dynamical
process in the course of which the space-time passes {\em the
"degenerate" phase where both the metrical structure and the total
4D-volume measure disappear}. The latter means also that the term
"orientation of the space-time manifold" loses any sense at
moments $t_i$ ($i=1,2,3,...$). We conclude therefore that two 4D
differentiable manifolds with opposite orientations (described by
means of a sign indefinite volume 4-form) equipped with connection
and metrical structure   still are not enough to describe the
arena of the gravitational dynamics: the complete description of
the space-time dynamics requires also the mentioned degenerate
phase. This situation is somewhat similar to that  discussed in
Introduction: first-order formulation of GR where the degenerate
phase with $g_{\mu\nu}= 0$ should be also
added\cite{Hawking1979},\cite{D'Auria-Regge},\cite{Tseytlin1982};
see e.g. the recent discussion by Ba\~{n}ados\cite{Banados} where
the limiting process $g_{\mu\nu}\to 0$ is analyzed.

 A new interesting feature of ground states in TMT we have
 revealed in the present paper concerns the so-called global Einstein symmetry (GES),
 Eqs.(\ref{LES1})-(\ref{linear-trans}), which turns out generically to be explicitly broken
 in all models with non-trivial dynamics. The surprising result we
 have discovered here on the basis of a number of models
 is that {\em the GES is restored in the course of
 transitions to the ground state} in all models considered.
 Hence its subgroup
of the sign inversions of $g_{\mu\nu}$ and $\Phi$,
Eq.(\ref{reflection}), is also restored. Therefore the
oscillations of $g_{\mu\nu}$ and $\Phi$ around zero in the course
of transition to a  $\Lambda =0$ ground state provoke a wish to
compare this {\em dynamical effect} with the attempts to solve the
old CC problem developed in Refs.\cite{Erdem}-\cite{reflection2}.
The main idea of these approaches is that the field theory or at
least the ground state\cite{'t Hooft}  should be invariant under
transformations of a discrete symmetry. According to
Refs.\cite{Erdem}-\cite{reflection2} it might be either an
invariance under the metric reversal symmetry or under the
space-time coordinate transformations with the imaginary unit $i$:
$x^A\rightarrow ix^A$. In contrast with these approaches, in TMT
there is no need to postulate such exotic enough symmetries.
Nevertheless we have seen that sign inversions of $g_{\mu\nu}$
emerge as a dynamical effect in the course of the cosmological
evolution and this effect has indeed a relation to the resolution
of the old CC problem.

\subsubsection{The case of a tiny CC}

 In the scalar field models of dark energy, an
interesting feature of TMT consists in a possibility to provide a
small value of the CC. If in the model of Sec.VIII.B, the
parameter $V_2<0$ and $|V_2|\gg b_gV_1$ then the CC can be very
small without the need for $V_1$ and $V_2$ to be very small. For
example, if $V_{1}$ is determined by the
 energy scale of electroweak symmetry breaking $V_{1}\sim
(10^{3}GeV)^{4}$ and $V_{2}$ is determined by the Planck scale
$V_{2} \sim (10^{18}GeV)^{4}$ then $\Lambda_{1}\sim
(10^{-3}eV)^{4}$. Along with such a seesaw  mechanism\cite{G1},
\cite{seesaw}, there exists another way to explain the smallness
of the CC applicable in all types of scenarios discussed in
Secs.VIII.B-VIII.D (see also Appendix A). As one can see from
Eqs.(\ref{lambda}), (\ref{a-phi-exact-bV1equal2V2}) and
(\ref{minVeff}), the value of $\Lambda$ appears to be inverse
proportional\footnote{In the pure gravity model, Sec.III,
$\Lambda$ is proportional to $b_g$} to the dimensionless parameter
$b_g$ which characterizes the relative strength of the 'manifold'
and 'metrical' parts of the gravitational action. If for example
$V_{1}\sim (10^{3}GeV)^{4}$ then for getting $\Lambda_1\sim
(10^{-3}eV)^{4}$ one should assume that $b_{g}\sim 10^{60}$. Such
a large value of $b_{g}$ (see Eq.(\ref{L1L2})) permits to
formulate  {\em a correspondence principle}\cite{GK9} {\em between
TMT and regular (i.e. one-measure) field theories}: when
$\zeta/b_g\ll 1$ then one can neglect the gravitational term in
$L_1$ with respect to that in $L_2$ (see Eq.(\ref{L1L2}) or
Eq.(\ref{S-model-scalar.f.}) or Eq.(\ref{totaction})). More
detailed analysis shows that in such a case the manifold volume
measure $\Phi =\zeta \sqrt{-g}$ has no a dynamical effect and TMT
is reduced to GR.  This happens e.g. in the model of Sec.VIII.C
where the late time evolution proceeds in a quintessence-like
manner: the energy density decreases to the cosmological constant,
Eq.(\ref{Lambda-bV1equal2V2}), and $\zeta\to 0$,
Eq.(\ref{zeta-asympt-delta=0-1}). Another example is the model of
Sec.VIII.D where $\Phi = 0$ in the ground state,
Eq.(\ref{zeta-in-minVeff}), while $\sqrt{-g}$ is finite. However
generically $\zeta/b_g$ is not small, as it happens for example in
the quintessence-like scenario of the late time universe in the
model of Sec.VIII.B (see Eq.(\ref{zeta-asympt-delta=0})).

\subsection{Possibilities for predictions of new physical effects}

\subsubsection{Short review of the TMT model with spontaneously broken
dilatation symmetry in the presence of matter}

It would be interesting to find out other possible physical
manifestations of the  sign indefiniteness of the manifold volume
measure. In fact, the model with spontaneously broken dilatation
symmetry studied in Secs.VII-IX and in Ref.\cite{GK9} allows
extensions which include fermion and gauge
fields\cite{GK5}-\cite{GK7} or, alternatively, dust as a
phenomenological matter model\cite{GK10}. In the former case, for
example, the constraint (\ref{constraint2}) is modified to the
following
\begin{equation}
\frac{1}{(\zeta
+b)^{2}}\left\{(b-\zeta)\left[M^{4}e^{-2\alpha\phi/M_{p}}+
V_{1}\right]-2V_{2})\right\}= \frac{\mu(\zeta -\zeta_{1})(\zeta
-\zeta_{2})}{2(\zeta +k)^2(\zeta +b)^{1/2}}\bar{\Psi}\Psi,
\label{constraint3}
\end{equation}
where $\Psi$ is the fermion field in the Einstein frame and for
simplicity we have chosen $\delta =0$ (that is $b_{\phi}=b_g=b$);
$\zeta_{1,2}$ are defined by
\begin{equation}
\zeta_{1,2}=\frac{1}{2}\left[k-3h\pm\sqrt{(k-3h)^{2}+ 8b(k-h)
-4kh}\,\right].
 \label{zeta12}
\end{equation}
and the dimensionless parameters $k$ and $h$ appear in the
underlying action in the total volume measures of the fermion
kinetic term
\begin{equation}
\int d^{4}x e^{\alpha\phi /M_{p}}(\Phi +k\sqrt{-g})
\frac{i}{2}\overline{\Psi}
\left(\gamma^{a}e_{a}^{\mu}\overrightarrow{\nabla}_{\mu}-
\overleftarrow{\nabla}_{\mu}\gamma^{a}e_{a}^{\mu}\right)\Psi
 \label{k}
\end{equation}
 and the fermion mass term
\begin{equation}
 -\int d^{4}xe^{\frac{3}{2}\alpha\phi /M_{p}}
(\Phi +h\sqrt{-g})\mu\overline{\Psi}\Psi
 \label{h}
\end{equation}
 respectively. Note that the fermion equation in the Einstein frame has
  a canonical form but the mass of the fermion turns out $\zeta$
  dependent
\begin{equation}
m(\zeta)= \frac{\mu(\zeta +h)}{(\zeta +k)(\zeta +b)^{1/2}}
\label{fermion_mass}
\end{equation}

   The constraint (\ref{constraint3}) describes the
local balance between the fermion energy density and the scalar
field $\phi$ contribution to the dark energy density in the
space-time region where the wave function of the primordial
fermion is not equal to zero. By means of this balance the
constraint determines the scalar $\zeta(x)$.

In the case of dust as a phenomenological matter model, the r.h.s.
of the constraint (\ref{constraint3}) looks
\begin{equation}
\frac{\zeta -b_m +2b}{2\sqrt{\zeta +b}}\, m\, \tilde{n},
\label{rhs-dust}
\end{equation}
where the dimensionless parameter $b_m$ appears in the total
volume measure of the dust contribution to the underlying action
\begin{equation}
S_{m}=\int (\Phi +b_{m}\sqrt{-g})L_m d^{4}x \label{dust}
\end{equation}
\begin{equation}
L_m=-m\sum_{i}\int e^{\frac{1}{2}\alpha\phi/M_{p}}
\sqrt{g_{\alpha\beta}\frac{dx_i^{\alpha}}{d\lambda}\frac{dx_i^{\beta}}{d\lambda}}\,
\frac{\delta^4(x-x_i(\lambda))}{\sqrt{-g}}d\lambda \label{Lm}
\end{equation}
and $m$ is the mass parameter.

 The wonderful feature of these models in the Einstein frame
 consists of the exact coincidence of the following three quantities:
 a) the noncanonical (in comparison with GR) terms in the
 energy-momentum tensor; b) the effective coupling "constant" of the
dilaton $\phi$ to the matter (up to the factor $\alpha/M_p$); c)
the expressions in the r.h.s. of the above mentioned constraints
(\ref{constraint3}) and (\ref{rhs-dust}) for fermionic matter and
dust respectively. For {\em matter in normal conditions}, the
local matter energy density (i.e. in the space-time region
occupied by the matter) is many orders of magnitude larger than
the vacuum energy density. Detailed
analysis\cite{GK5}-\cite{GK7},\cite{GK10} shows that when the
matter is in the normal conditions,  the balance dictated by the
constraint becomes possible if $\zeta$ with very high accuracy
takes the constant values: $\zeta\approx\zeta_1$ or
$\zeta\approx\zeta_2$ for fermions (and therefore the fermion
masses become constant)) and $\zeta\approx b_m-2b$ for the dust.
Then the mentioned three quantities simultaneously become
extremely small. Besides for the matter in normal conditions the
gravitational equations are reduced to the canonical GR equations.
The practical disappearance of the dilaton-to-matter coupling
"constant" for the matter in normal conditions which occurs
without fine tuning of the parameters allows us to assert that in
such type of models the fifth force problem is
resolved\cite{GK7},\cite{GK10}.

  It does not mean however
 that matter does not interact with the dilaton at all. When the matter
 is in  states different from normal, the effect of
 dilaton-to-matter coupling may yield new very interesting
 phenomena. One of such effects  appears when the neutrino energy
  density decreases to the order of magnitude close to the vacuum
  energy density. The latter can happen due to spreading of the neutrino
  wave packet. Then the cold gas of uniformly distributed nonrelativistic
 neutrinos causes a reconstruction of the vacuum to a state with $\zeta\to
 |k|$ and as a result the neutrino gas rapidly transmute into an
 exotic state called neutrino dark energy(see e.g. Ref.\cite{Nelson}).
 This effect was studied in details
  in Ref.\cite{GK7} where we have shown that transmutation from
  the pure scalar field dark energy to the neutrino dark energy regime is
favorable from the energetic point of view.

\subsubsection{Prediction of strong gravity effect in high energy physics experiments}

For the solutions $\zeta\approx\zeta_1$ or $\zeta\approx\zeta_2$
of the constraint (\ref{constraint3}), the l.h.s. of the
constraint has the order of magnitude close to the vacuum energy
density. There exists however another solution if one  allows a
possibility that in the core of the support of the fermion wave
function the local dark energy density may be much bigger than the
vacuum energy density. Such a solution turns out to be possible as
fermion density is very big and $\zeta$ becomes negative and close
enough to the value $\zeta\approx -b$. Then the solution of the
constraint (\ref{constraint3}) looks\cite{GK5}
\begin{equation}
\frac{1}{\sqrt{\zeta
+b}}\approx\left[\frac{\mu(b-h)}{4M^4b(b-k)}\bar{\Psi}\Psi
e^{2\alpha\phi/M_p}\right]^{1/3}.
 \label{zeta-b}
\end{equation}
In such a case, instead of constant masses, as it was for
$\zeta\approx\zeta_{1,2}$, Eq.(\ref{fermion_mass}) results in the
following fermion self-interaction term in the effective fermion
Lagrangian
\begin{equation}
L^{ferm}_{selfint}=3\left[\frac{1}{b}\left(\frac{\mu(b-h)}{4M(b-k)}\bar{\Psi}\Psi\right)^4
e^{2\alpha\phi/M_p}\right]^{1/3}.
 \label{selfinter}
\end{equation}

It is very interesting that the described effect is the direct
consequence of the strong gravity. In fact, in the regime where
$\zeta +b\ll 1$  the effective Newton constant in the
gravitational term of underlying action(\ref{totaction})
\begin{equation}
S_{grav}=-\int d^{4}x \sqrt{-g}\,\frac{\zeta +b}{\kappa b}
R(\Gamma ,g)e^{\alpha\phi /M_{p}}
 \label{strong-grav}
\end{equation}
becomes anomalously  large. Recall that for simplicity we have
chosen here $b_{\phi}=b_g=b$. But if one do not to imply this fine
tuning then one can immediately see from
Eqs.(\ref{gef1})-(\ref{Veff2}) that in the Einstein frame the
regime of the strong gravity dictated by the dense fermion matter
is manifested for the dilaton too.

 The coupling constant in Eq.(\ref{selfinter}) is dimensionless
and depends exponentially of the dilaton $\phi$ if one can regard
$\phi$ as a background field $\phi=\bar{\phi}$. But in a more
general case Eq.(\ref{selfinter}) may be treated as describing an
anomalous dilaton-to-fermion interaction very much different from
the discussed above case of interaction of the dilaton to the
fermion matter in normal conditions where the coupling constant
practically vanishes. Such an anomalous dilaton-to-fermion
interaction should result in  creation of quanta of the dilaton
field in processes with very heavy fermions.  The probability of
these processes is of course proportional to the Newton constant
$M_p^{-2}$. But the new effect consists of the fact that the
effective coupling constant of the anomalous dilaton-to-fermion
interaction is proportional to $e^{2\alpha\bar{\phi}/3M_p}$. If
the dilaton is the scalar field responsible for the quintessential
inflation type of the cosmological scenario\cite{Quint-ess} then
one should expect an exponential amplification of the effective
coupling of this interaction in the present day universe in
comparison with the early universe. One can hope that the
described effect of the strong gravity
 might be revealed  in the LHC experiments in the form of missing energy
 due to the multiple production of quanta of the dilaton
field (recall that coupling of the dilaton to fermions in normal
conditions practically vanishes and therefore the dilaton will not
be observed after  being emitted).

\subsubsection{Some other possible effects}

1. {\em Dark matter as effect of gravitational enhancement.} In
the case of dust as a phenomenological matter model, the
constraint (\ref{constraint3}) with the r.h.s. (\ref{rhs-dust}) is
the fifth degree algebraic equation with respect to $\sqrt{\zeta
+b}$. There are some indications that in a certain region of the
parameters a solution of the constraint exists which could provide
a very interesting effect of an  amplification of the
gravitational field of visible diluted galactic and intergalactic
dust or/and neutrinos. Such an effect might imply that the dark
matter is not a new sort of matter but it is just a result of a so
far unknown enhancement of the gravitational field of  low density
states of usual matter.

2. {\em Dilaton to photon coupling.} Astrophysical observations of
few last  years indicate anomalously large transparency of the
Universe to gamma rays\cite{Nature1},\cite{Nature2}. It is hard to
explain this astrophysical puzzle in the framework of
extragalactic background light. Recently a natural mechanism was
suggested by De Angelis, Mansutti and Roncadelli\cite{Angelis} in
order to resolve this puzzle. The idea is to suppose that there
exists a very light spin-zero boson  coupled to the photon:
\begin{equation}
L_{\phi\gamma}=-\frac{1}{4\mu}F_{\mu\nu}\tilde{F}_{\mu\nu}\phi.
 \label{phi-to-photon}
\end{equation}
 where $\mu$ is a mass parameter. In
the context of quintessential scenario such a coupling was studied
by Carroll\cite{carroll}. Then
$\gamma\longrightarrow\phi\longrightarrow\gamma$ oscillations
emerge which explain\cite{Angelis}  the observed transparency of
the Universe to gamma rays in a natural way if mass of the
spin-zero boson $m<10^{-10}eV$. The crucial feature of this boson
is that no other coupling of this scalar to matter exists. In the
standard quintessence models this feature seems to be a real
problem. But in TMT, as we already mentioned (see also
Refs.\cite{GK7},\cite{GK10}) the dilaton playing the role of
quintessence field decouples from matter in normal conditions. At
the same time its coupling to the photon in the form
(\ref{phi-to-photon}) is not suppressed.

3. {\em Creation of a universe in the laboratory.}  A theoretical
attempt by Farhi, Guth and Guven to describe a creation of a
universe in the laboratory\cite{Guth} runs across a need to allow
vanishing and changing sign of $\sqrt{-g}$. In Ref.\cite{Guth},
this need is naturally regarded as a pathology. If similar
approach to the  problem of creation of a universe in the
laboratory could be formulated in the framework of TMT then
instead of $\sqrt{-g}$ there should appear a linear combination of
$\Phi$ and $\sqrt{-g}$ which, as we already know, is able to
vanish and change sign. In recent paper\cite{GS} by Guendelman and
Sakai a model of child universe production without initial
singularities was studied. To provide the desirable absence of
initial singularity a crucial point is that the energy momentum
tensor of the domain wall should be dominated by a sort of phantom
energy. A possible way to realize this idea is to apply the
dynamical  brane tension\cite{GKNP} obtained when using the
modified volume measure similar to the signed measure $\Phi$ of
the present paper. So it could be that applying the notions
explored in the present paper one can obtain also a framework for
formulating non singular child universe production.

4. {\em Unparticle physics.} $\zeta$ dependence of the fermion
mass, Eq.(\ref{fermion_mass}), together with the constraint
(\ref{constraint3}) can be treated as a $\bar{\Psi}\Psi$
dependence of the fermion mass. This means that in states
different from the normal one, the fermion mass spectrum may be
continuous, that allows to think of a possibility to establish
relation with the idea of unparticle physics\cite{Georgi}.

Note finally that for the matter in normal conditions the model
does not impose   essential constraints on the parameters of the
model (such as $b_g$, $b_{\phi}$, $b_m$, $k$, $h$). But the
appropriate constraints should appear when more progress in the
study of the listed and another possible new effects will be
achieved.

\section{Acknowledgements}

We acknowledge V. Goldstein, M. Lin, E. Nissimov and S. Pacheva
for useful discussions of some mathematical subjects. We also
thank M. Duff for explaining us his approach and M. Ba\~{n}ados
for useful conversations. We are also grateful to the referee
whose constructive remarks assisted us in the improvement of this
paper.

\appendix

\section{The Ground State with Non Zero CC in Model I }

Let us consider the scalar field model I
(Eqs.(\ref{S-model-scalar.f.}) and (\ref{V12model}))
 with $\delta =0$ where we now
choose a positive integration constant ($s=+1$) and   the
parameters $V_2^{(0)}<0$, $b_g\mu_1^2>\mu_2^2$. Then the ground
state is realized for $\phi =0$ and the vacuum energy is
\begin{equation}
\Lambda =V_{eff}(0)=\frac{M^8}{4b_gM^4-V_2^{(0)}} \label{AppI}
\end{equation}
In this ground state, both the measure $\Phi>0$ and all components
of the metric tensor $g_{\mu\nu}$ are regular. Note that the
presence of the free dimensionless parameter $b_g$ in the
denominator allows again to reach a small vacuum energy by means
of the correspondence principle discussed in item 2 of Sec.X.

\section{Global Einstein Symmetry does not Guarantee Resolution of
the CC Problem}

 In the model (\ref{S+Delta}), the gravitational equations
are modified to the following
\begin{equation}
G_{\mu\nu}(\tilde{g})=\frac{\kappa}{2}\left[\phi_{,\mu}\phi_{,\nu}-\frac{1}{2}\tilde{g}_{\mu\nu}X
+\frac{b_g[sM^4+V_1(\phi)+2\lambda\zeta]-V_2(\phi)-\lambda\zeta^2}{(\zeta
+b_g)^2}\right] \label{grav-App}
\end{equation}
while the form of the scalar field $\phi$ equation remains the
same as in Eq.(\ref{phief}). However the constraint is now very
much differs from Eq.(\ref{constraint2-1}):
\begin{equation}
4\lambda\zeta^2 +[sM^4+V_1(\phi)-2b_g\lambda]\zeta +2V_2(\phi)
-b_g[sM^4+V_1(\phi)]=0\label{constr-App}
\end{equation}
One can see from Eq.(\ref{grav-App}) that $\zeta$-dependence
emerges now in the numerator of the effective potential. Besides,
it is evident that in contrast with what was in Sec.V, the regime
with $\zeta\to\infty$ cannot be a solution of the constraint. It
is evident that  a zero minimum of the effective potential cannot
be now reached without fine tuning. Thus although the second term
in the action (\ref{S+Delta}) is invariant under the GES, adding
this term we loss  the ability to resolve the old CC problem.


\begin{thebibliography}{99}
\bibitem{Einstein}
Einstein A and Rosen N 1935 {\it Phys. Rev.} \textbf{48} 73

\bibitem{Hawking1979}
 Hawking S W {\it Nucl.Phys.} 1978 B{\it 144} 349; Hawking S
W 1979 in {\em Recent Developments in Gravitation} ed M Levy and S
Deser (New York; Plenum)

\bibitem{D'Auria-Regge}
D'Auria R and Regge T 1982 {\em Nucl. Phys.} B \textbf{195} 308

\bibitem{Tseytlin1982}
Tseytlin A A 1982 {\em J. Phys. A: Math. Gen.} \textbf{15} L105

\bibitem{Horowitz}
Horowitz Gary T 1991 Class Quantum Grav. \textbf{8} 587

\bibitem{Ashtekar-3-in-Jacobson}
Ashtekar A 1991 {\em Lectures on Non-Perturbative Canonical
Gravity} (World Scientific)

\bibitem{Jacobson-2-in-Jacobson}
Jacobson T and Smolin L 1988 {\em Nucl. Phys.} B \textbf{299} 295

\bibitem{Dray-0}
Dray T,  Manogue C A and  Tucker R W 1991 {\em Gen. Rel. Grav.}
{\bf 23} 967

\bibitem{Ellis}
Ellis G, Sumeruk A, Coule D and Hellaby C 1992 {\em Class. Quantum
Grav.} \textbf{9} 1535

\bibitem{Odintsov}
 Elizalde E,
Odintsov S and  Romeo A 1994 {\em Class. Quant. Grav.} \textbf{11}
61

\bibitem{Dray-1}
 Dray T,  Ellis G,
Hellaby C and Corinne A. Manogue C A 1997 {\em Gen. Rel. Grav.}
{\bf 29} 591

\bibitem{Dray-2}
Dray T,  Ellis G, Hellaby C 2001 {\em Gen. Rel. Grav.} {\bf 33}
1041


\bibitem{Volovich}
  Borowiec A,  Francaviglia M and  Volovich I 2007.
 {\em Int. J. Geom. Meth. Mod. Phys.} \textbf{4} 647

\bibitem{Senovilla}
 Mars M,  Senovilla Jose M M and  Vera R 2008 {\em Phys. Rev.} D \textbf{77} 027501

\bibitem{Witten-15-and-16-in-Horowitz}
Witten E 1988 {\em Commun. Math. Phys.} \textbf{117} 353\\
Witten E 1988 {\em Nucl. Phys.} B \textbf{311} 46

\bibitem{Giddings}
Giddings S B 1991 Physics Letters B \textbf{268} 17

\bibitem{Banados}
Ba\~{n}ados M  2007 {\em Class. Quantum Grav.}\textbf{24} 5911\\
Ba\~{n}ados M  2008 {\em Phys.Rev.} \textbf{D} 77
 123534

\bibitem{signed}  Cohn D L, {\it Measure
Theory}, Birkhauser, Boston, 1993.


\bibitem{Taylor}
Taylor J G 1979 {\em Phys. Rev.} \textbf{19} 2336

\bibitem{Wilczek}
Wilczek F 1998 {\em Phys. Rev. Lett.} \textbf{80} 4851

\bibitem{Mosna}
Mosna R A and Saa A 2005 {\em J.Math.Phys.} \textbf{46} 112502

\bibitem{GK1}
Guendelman E I and Kaganovich A B  {\em Phys. Rev.} 1996
D\textbf{53} 7020;  {\em Mod. Phys. Lett.} 1997  A\textbf{12}
2421; {\em Phys. Rev.} D\textbf{55} 5970;   {\em Mod. Phys. Lett.}
1997 A\textbf{12} 2421; {\em Phys. Rev.} 1997 D\textbf{56} 3548;
{\em Mod. Phys. Lett.} 1998 A\textbf{13} 1583.

\bibitem{GK2}
Guendelman E I and Kaganovich A B  {\em Phys. Rev.} 1998
D\textbf{57} 7200).

\bibitem{GK3}
Guendelman E I and Kaganovich A B {\em Phys. Rev.} 1999
\textbf{D60} 065004.

\bibitem{G1}
 Guendelman E I 1999 {\em Mod. Phys. Lett.} A{\bf 14}, 1043;
 {\it Class. Quant. Grav.} 2000 {\bf 17} 361;
gr-qc/9906025; {\it Mod. Phys. Lett.}  1999 A{\bf A4}, 1397;
gr-qc/9901067; hep-th/0106085; {\it Found. Phys.} 2001 {\bf 31}
1019;

\bibitem{K}
Kaganovich A B 2001 {\it Phys. Rev.} {\bf D63}, 025022.

\bibitem{GKatz}
Guendelman E I and Katz O 2003 {\it Class. Quant. Grav.} {\bf 20}
1715

\bibitem{G}
Guendelman E I 1997 {\it Phys. Lett.} B{\bf 412} 42; Guendelman E
I 2003 gr-qc/0303048; Guendelman E I and Spallucci E 2003
hep-th/0311102.

\bibitem{GK5}
Guendelman E I and Kaganovich A B 2002 {\it Int. J. Mod. Phys.}
A{\bf 17} 417.


\bibitem{GK6}
Guendelman E I and Kaganovich A B 2002 {\it Mod. Phys. Lett.}
A{\bf A7} 1227 (2002).

\bibitem{GK7}
Guendelman E I and Kaganovich A B 2004  hep-th/0411188; {\it
Int.J.Mod.Phys.} 2006 A{\bf 21} 4373.

\bibitem{GK8}
Guendelman E I and Kaganovich A B 2006  in {\it Paris 2005, Albert
Einstein's century},  AIP Conf.Proc. 2006 861 875, Paris;
hep-th/0603229.

\bibitem{GK9}
Guendelman E I and Kaganovich A B 2007 {\it Phys.Rev.} D{\bf 75}
083505.

\bibitem{GK10}
Guendelman E I and Kaganovich A B 2008 {\it Annals Phys.}. {\bf
323} 866.

\bibitem{Comelli}
Comelli D arXiv:0704.1802 [gr-qc].

\bibitem{Weinberg1}
Weinberg S 1989 {\it Rev. Mod. Phys.} {\bf 61} 1

\bibitem{unimodular-1}
Unruh W G 1989 {\it Phys. Rev.} 1989 D{\it 40} 1048.

\bibitem{unimodular-2}
Ng Y Jack and  van Dam H 1991 {\it J. Math. Phys.} {\bf 32} 1337.

\bibitem{k-essence}
Chiba T, Okabe T and Yamaguchi M 2000 {\it Phys.Rev.} D{\bf 62}
023511; Armendariz-Picon C,  Mukhanov V and  Steinhardt P J 2000
{\em Phys. Rev. Lett.} {\bf 85} 4438; {\it Phys. Rev.} 2001 D{\bf
63} 103510; Chiba T 2002 {\em Phys.Rev.}  D{\bf 66} 063514.

\bibitem{k-inflation-Mukhanov}
 Armendariz-Picon C.,  Damour T and  Mukhanov V F 1999 {\em
 Phys.Lett.}
B{\bf 458} 209.

\bibitem{Phantom-usual}
  Caldwell R R {\em Phys.Lett.} 2002 B{\bf 545} 23;  Gibbons G W 2003,
hep-th/0302199.


\bibitem{Erdem}
 Erdem R 2005 {\em Phys. Lett.} B{\bf 621} 11; {\em Phys. Lett.} 2006 B{\bf
639} 348; {\em J. Phys.} 2007 A{\bf 40} 6945.

\bibitem{'t Hooft}
Nobbenhuis S 2006 {\em Found. Phys.} {\bf 36} 613; 't Hooft G,
Nobbenhuis S 2006 {\em Class. Quant. Grav.} {\bf 23} 3819.

\bibitem{reflection2}
 Duff M J and   Kalkkinen J. 2006 {\em Nucl. Phys.} B{\bf 758} 161;
2007 {\em Nucl. Phys.} B{\bf 760} 64.

\bibitem{seesaw}
  Arkani-Hamed N,  Hall L J,  Kolda C F and Murayama H 2000 {\em Phys.Rev.Lett.} {\bf 85} 4434.

\bibitem{Zee}
R. MacKenzie, F. Wilczek and A. Zee 1984 {\em Phys. Rev. Lett.}
{\bf 53} 2203.

\bibitem{Nelson}
R Fardon, A. E. Nelson, N. Weiner 2004 {\em JCAP} {\bf 0410} 005.

\bibitem{Quint-ess}
P.J.E. Peebles and A. Vilenkin 1999 {\em Phys.Rev} D{\bf 59}
063505.

\bibitem{Nature1}
Uchiyama Y, Aharonian F, Tanaka T, Takahashi T and Maeda 2007 {\em
Nature} {\bf 449} 576.

\bibitem{Nature2}
  Mazin D, Raue M 2007 {\em
Astron.Astrophys.} {\bf 471} 439.

\bibitem{Angelis}
 De Angelis A, Mansutti O
and Roncadelli M 2007 {\em Phys.Rev.} D{\bf 76} 121301.

\bibitem{carroll}
Carroll S 1998 {\it Phys. Rev. Lett.} {\bf 81} 3067.


\bibitem{Guth}
 Farhi E,  Guth A H  and  Guven J 1990
 {\em Nucl. Phys.} B{\bf 339} 417.

\bibitem{GS}
 Guendelman E I and Sakai N  2008 {\em Phys. Rev.} D{\bf 77}
 125002.

\bibitem{GKNP}
Guendelman E I, Kaganovich A B, Nissimov E, Pacheva S 2002 {\em
Phys. Rev.} D{\bf 66} 046003; "Impact of dynamical tensions in
modified string and brane theories", Presented at 5th
International Workshop on Lie Theory and Its Applications in
Physics, Varna, Bulgaria, 16-22 Jun 2003, H.D. Doebner and V.
Dobrev Eds., World Scientific, 2004. Published in {\em Varna 2003,
Lie theory and its applications in physics V} 241-250 e-Print:
hep-th/0401083.

\bibitem{Georgi}
 Georgi H 2007 {\em Phys. Rev. Lett.} {\bf 98} 221601.


\end{thebibliography}
\end{document}